\documentclass{aa}

\usepackage{txfonts}
\usepackage{graphics}
\usepackage{epsfig}
\usepackage[totalwidth=17.8cm, totalheight=24.0cm]{geometry}
\usepackage{longtable}

\begin{document}

\title{The VMC Survey - VI. Quasars behind the Magellanic system\thanks{Based on observations made with VISTA at the Paranal Observatory
under program ID 179.B-2003.}}

\author{M.-R.L. Cioni\inst{1, 2,}\thanks{Research Fellow of the Alexander von Humboldt Foundation}
	\and D. Kamath\inst{3}
	\and S. Rubele\inst{4}
	\and J.Th. van Loon\inst{5}
	\and P.R. Wood\inst{3}
	\and J.P. Emerson\inst{6}
	\and B. K. Gibson\inst{7}
	\and M.A.T. Groenewegen\inst{8}
	\and V.D. Ivanov\inst{9}
	\and B. Miszalski\inst{10,11}
	\and V. Ripepi\inst{12}	
}

\offprints{mcioni@usm.uni-muenchen.de}

\institute{
	University Observatory Munich, Scheinerstrasse 1, 81679 M\"{u}nchen, Germany
	\and University of Hertfordshire, Physics Astronomy and Mathematics, Hatfield AL10 9AB, United Kingdom
	\and RSAA, Mount Stromlo Observatory, Cotter Road, Weston Creek, ACT 2611, Australia	
	\and INAF, Osservatorio Astronomico di Padova, Vicolo dell'Osservatorio 5, 35122 Padova, Italy
	\and Keele University, Lennard-Jones Laboratories, ST5 5BG, United Kingdom
	\and School of Physics and Astronomy, Queen Mary University of London, Mile End Road, London E1 4NS, United Kingdom
	\and Jeremiah Horrocks Institute, University of Central Lancshire, Preston PR1 2HE, United Kingdom
	\and Royal Observatory of Belgium, Ringlaan 3, 1180 Ukkel, Belgium
	\and European Southern Observatory, Av. Alonso de C\'{o}rdoba 3107, Casilla 19, Santiago, Chile
	\and South African Astronomical Observatory, PO Box 9, Observatory, 7935, South Africa
	\and Southern African Large Telescope Foundation, PO Box 9, Observatory, 7935, South Africa
	\and INAF, Osservatorio Astornomico di Capodimonte, via Moiariello 16, 80131 Napoli, Italy
}

\date{Received 28 May 2012 / Accepted 9 October 2012}

\titlerunning{QSOs behind the Magellanic system}

\authorrunning{Cioni et al.}

\abstract{The number and spatial distribution of confirmed quasi-stellar objects (QSOs) behind the Magellanic system is limited. This undermines their use as astrometric reference objects for different types of studies.}
{We have searched for criteria to identify candidate QSOs using observations from the VISTA survey of the Magellanic Clouds system (VMC) that provides photometry in the $YJK_\mathrm{s}$ bands and $12$ epochs in the $K_\mathrm{s}$ band.}
{The $(Y-J)$ versus $(J-K_\mathrm{s})$ diagram has been used to distinguish QSO candidates from Milky Way stars and stars of the Magellanic Clouds. Then, the slope of variation in the $K_\mathrm{s}$ band has been used to identify a sample of high confidence candidates. These criteria were developed based on the properties of $117$ known QSOs presently observed by the VMC survey.}
{VMC $YJK_\mathrm{s}$ magnitudes and $K_\mathrm{s}$ light-curves of known QSOs behind the Magellanic system are presented. About $75$\% of them show a slope of variation in K$_\mathrm{s}$ $>10^{-4}$ mag/day and the shape of the light-curve is in general irregular and without any clear periodicity.
The number of QSO candidates found in tiles including the South Ecliptic Pole and the 30 Doradus regions is $22$ and $26$, respectively, with a $\sim20$\% contamination by young stellar objects, planetary nebulae, stars and normal galaxies. }
{By extrapolating the number of QSO candidates to the entire VMC survey area we expect to find about $1\,200$ QSOs behind the LMC, $400$ behind the SMC, $200$ behind the Bridge and $30$ behind the Stream areas, but not all will be suitable for astrometry. Further, the $K_\mathrm{s}$ band light-curves can help support investigations of the mechanism responsible for the variations. }


\maketitle

\section{Introduction}
\label{intro}
The astrometric accuracy and the photometric sensitivity of observations made with VISTA is sufficiently good that we expect data from the VISTA Magellanic Clouds survey (VMC; Cioni et al. \cite{cio11} hereafter Paper I) can be used to derive proper motions of the Magellanic Clouds (MCs). Such proper motion studies require a reference grid of bright, distant, non-moving, point like objects. Quasi-stellar objects (QSOs) provide such a grid (e.g. Kallivayalil et al. \cite{kal06}; Costa et al. \cite{cos11}). QSOs are point-like sources believed to be powered by accretion onto black holes in the centre of distant galaxies. QSOs can also be used as background sources to examine the composition of the MC interstellar medium along the line of sight (e.g. Redfield et al. \cite{red06}; van Loon et al. \cite{loo09}), and are important for studies of galaxy formation and evolution.

The density of QSOs with $i<19$ mag is $\sim 11$ per deg$^2$ as estimated from the Sloan Digital Sky Survey (SDSS) Quasar Catalogue (Schneider et al. \cite{sch10}) but discovery of candidate QSOs behind the MCs is complicated by the necessity to distinguish candidate QSOÕs from the dense stellar content of the MCs themselves. Candidate QSOs must then be observed spectroscopically to confirm which are true QSOs, and to make this process efficient the sample of candidate QSOs should be as clean as possible. Selection of candidate QSOs behind the MCs has been greatly improved by long-term multi-epoch observations, as part of micro-lensing projects such as the MAssive Compact Halo Objects (MACHO -- Alcock et al. \cite{alc00}) and the Optical Gravitational Lensing Experiment (OGLE -- Udalski et al. \cite{uda92}), and using data at various wavelengths, from ultraviolet (UV) to infrared (IR), that permit better removal of stellar objects (young stellar objects, planetary nebulae, hot and red stars) from QSO candidate samples. Methods used to identifying candidate QSOs include X-ray (Shanks et al. \cite{sha91}) or radio emission (Schmidt \cite{sch68}), mid-IR (Stern et al. \cite{ste05}) and near-IR colours (see below). Flux variations probably associated with the accretion disc have also been used (Hook et al. \cite{hok94}). QSO candidates are spectroscopically confirmed on the basis of optical and UV ionic emission lines, from which their redshifts are measured (e.g. Vanden Berk et al. \cite{vbe01}). The most recent such investigation by Koz{\l}owski et al. (\cite{koz12}) focused on the southern edge of the Large Magellanic Cloud (LMC) and is relatively complete for objects with $I< 19.2$ mag, with a candidate sample based on Spitzer Space Telescope mid-IR colours, X-ray emission and/or optical variability. Spectra of their sample quadrupled the number of confirmed QSOs behind the LMC. 

Currently there are $360$ known QSOs behind the Magellanic system of which $233$ are behind the LMC, $100$ behind the Small Magellanic Cloud (SMC) and $27$ behind the inter-cloud region including the Magellanic Bridge, and many more QSO candidates awaiting follow-up observations to establish their nature (e.g. Koz{\l}owski \& Kochanek \cite{koz09}; Kim et al. \cite{kim12}). However these objects cover a limited area compared to the extent of the whole MC system being surveyed in the VMC survey. VMC detects sources as faint as $K_\mathrm{s}=23.4$ mag (AB) with S/N$=5$, corresponding to the luminosity of sources below the old main-sequence turnoff in the LMC which occurs at $I\sim 22$ mag. The $YJK_\mathrm{s}$ VMC survey, which is multi-epoch in $K_\mathrm{s}$, has the potential to considerably enlarge the parameter space for the search of QSOs, behind the Magellanic system, both in terms of sensitivity and spatial distribution. Near-IR criteria to select QSOs have been proposed previously. Kouzuma \& Yamaoka's (\cite{kou10}) criteria were based on 2MASS $JHK$ bands. A series of papers have described the $K$ excess ($K$X) method (Warren, Hewett \& Foltz \cite{war00}) using UKIDSS $JK$ bands and the SDSS $g$ band (e.g. Maddox et al. \cite{mad12}; Mortlock et al. \cite{mor12}). Findlay et al. (\cite{fin12}) based their selection on single epoch VISTA $ZYJ$ data.

Our aim is to establish VMC $YJK_\mathrm{s}$ bands selection criteria using the known QSOs behind the Magellanic system which have already been observed, and investigate their utility in identifying new candidate QSOs in the areas being surveyed. The first stage of our method uses the $(Y-J)$ vs $(J-K_\mathrm{s})$ diagram and the second stage uses $K_\mathrm{s}$ variability.

Section \ref{data} describes the VMC data for the known QSOs behind the Magellanic system that have been covered so far. Section \ref{results} uses VMCÕs $YJK_\mathrm{s}$ photometry of these QSOs to establish colour criteria for isolating them from Milky Way (MW) and MC objects, and to examine the $K_\mathrm{s}$ variability of the QSOs over a baseline of up to $12$ epochs over two years. Section \ref{discussion} discusses the reliability of our method and Sect. \ref{conclusions} presents our conclusions. The Appendix provides further information on the known QSOs and their $K_\mathrm{s}$ band light-curves. The influence of reddening and of photometrically selected non-QSOs on our selection criteria is also discussed in there.

\section{Data and Sample}
\label{data}

\subsection{VMC data}
\label{data1}
The near-IR data analysed in this study were obtained with the Visual and Infrared Survey Telescope for Astronomy (VISTA; Emerson \& Sutherland \cite{eme10}) for the VMC survey and include observations acquired until the end of September 2011. The data were reduced onto the VISTA photometric system (Vegamag $=0$) with the VISTA Data Flow System (VDFS) pipeline v1.1 (Irwin et al. \cite{irw04}) and extracted from the VISTA Science Archive\footnote{http://horus.roe.ac.uk//login.html} (VSA).
The survey strategy involves repeated observations of tiles, where one tile homogeneously covers an area of $1.5$ deg$^2$ with $3$ epochs in both  $Y$ and $J$ filters, and $12$ epochs in $K_\mathrm{s}$ with the epochs spread over a time range of a year or longer. Details about the observing strategy and the data reduction are given in Paper I. 

Table \ref{vmcdata} lists the tiles that contain known QSOs and for which observations were obtained by end of September 2011. Tile names are specified in column $1$ and the limiting magnitude corresponding to sources with photometric errors $<0.1$ mag in columns $2-4$ while the number of available $K_\mathrm{s}$ epochs is given in column $5$ (see also Sect. \ref{secvar}). 

VMC sources have numerical quality flags and in the following analysis we discriminate sources that are well detected (Flag A - with quality flags $= 0-16$ where $16$ indicates that a source has been de-blended in at least one wave band), and those with a low quality detection (Flag B - with quality flags $>16$ that would arise if a source is located in the under-exposed edge areas of a tile, or in the upper half of detector \#16 which is known to have a varying quantum efficiency, that may result in bad flat fielding and unreliable magnitudes, or just a low confidence in the aperture magnitude of sources). 

\begin{table}
\caption{VMC tile sensitivity and epochs.}
\label{vmcdata}
\[
\begin{array}{ccccr}
\hline
\noalign{\smallskip}
 & \multicolumn{3}{c}{\mathrm{Limiting\,Magnitude}^a} & \mathrm{Epochs} \\
\mathrm{Tile} & Y & J & K_\mathrm{s} & K_\mathrm{s} \\
  &  \mathrm{(mag)} & \mathrm{(mag)} & \mathrm{(mag)} & \\
\hline
\noalign{\smallskip}
\mathrm{SMC}\,3\_3 & 20.659 & 20.412 & 19.607 & 9\\
\mathrm{SMC}\,3\_5 & 21.127 & 21.002 & 19.693 &  10\\
\mathrm{SMC}\,4\_3 & 20.152 & 19.977 & 19.071 &  3\\
\mathrm{SMC}\,5\_4 & 20.678 & 20.461 & 19.621 &  11\\
\mathrm{BRI}\,2\_3 & 21.222 & 20.952 & 19.734 &  9\\
\mathrm{BRI}\,3\_7 & 20.972 & 20.742 & 18.418 &  1\\
\mathrm{LMC}\,4\_2 & 21.305 & 20.593 & 19.501 &  8\\
\mathrm{LMC}\,4\_6 & 20.652 & 20.378 & 18.587 &  2\\
\mathrm{LMC}\,5\_5 & 19.890 & 19.626 & 19.081 &  10\\
\mathrm{LMC}\,6\_4 & 19.259 & 18.962 & 18.576 &  10\\
\mathrm{LMC}\,6\_6^b & 19.906 & 19.514 & 18.995 &  14\\
\mathrm{LMC}\,6\_8 & 20.782 & 20.585 & 17.911 & 1 \\
\mathrm{LMC}\,7\_3 & 20.406 & 19.949 & 18.802 &  2\\
\mathrm{LMC}\,8\_3 & 20.882 & 20.364 & 19.676 &  12\\
\mathrm{LMC}\,8\_8^b & 21.091 & 20.760 & 19.856 &  14\\
\mathrm{LMC}\,9\_7 & 20.989 & 20.414 & 18.226 &  1\\
\noalign{\smallskip}
\hline
\end{array}
\]

$^a$ For sources with photometric error $<0.1$ mag.

$^b$ VMC observations are completed.
\end{table}

\subsection{Known quasars}

\begin{figure*}
\resizebox{\hsize}{!}{\includegraphics{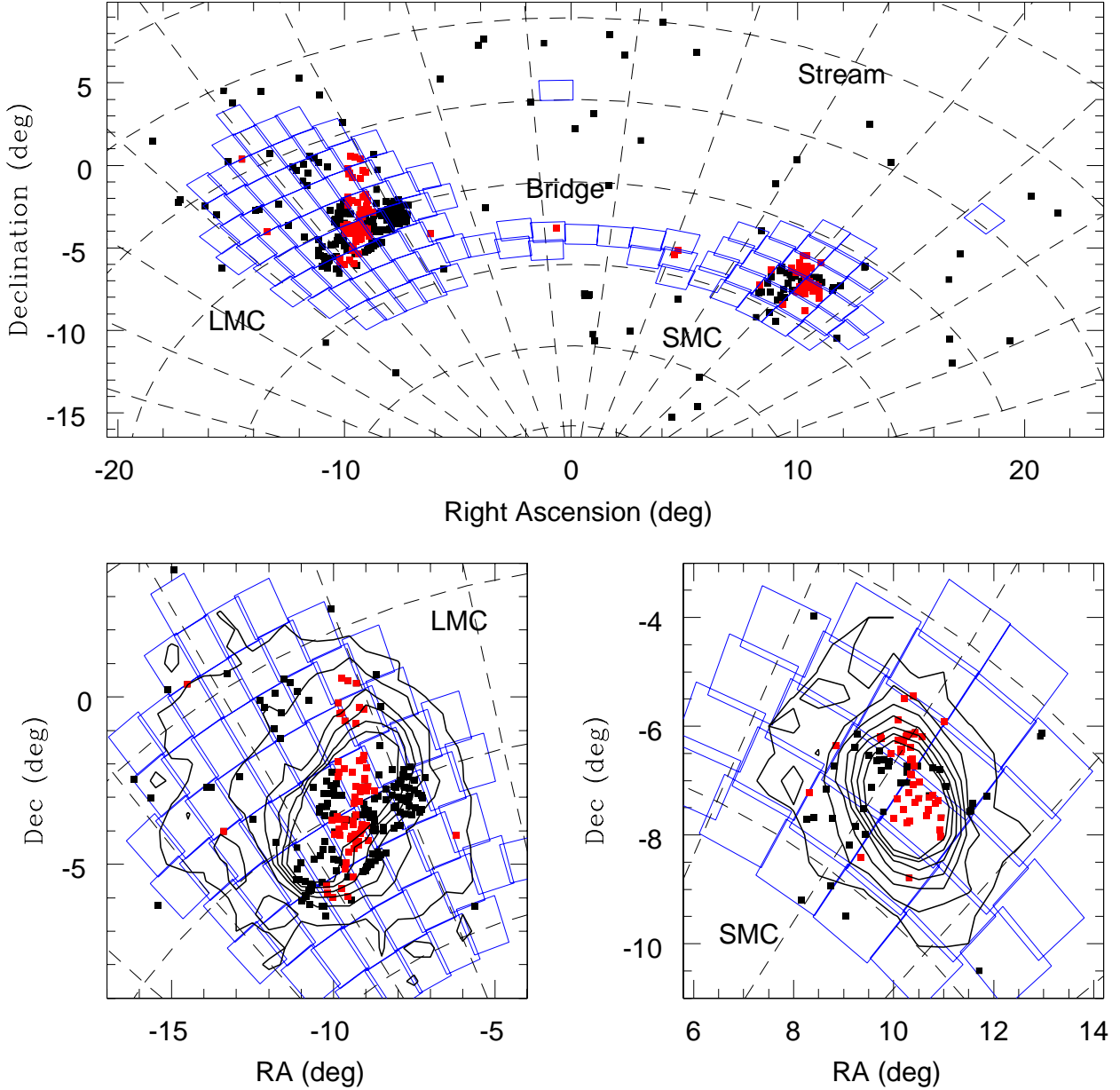}}
\caption{Distribution of known quasars behind the Magellanic system (top) with enlargements on the LMC (bottom-left) and SMC (bottom-right). All known quasars are shown with filled small black squares while those positionally matched with presently available VMC data are shown in red. Large empty blue rectangles indicate the VMC tiles. Contour plots show the distribution of evolved giant stars which delineate the outer structure and bar of the galaxies. Coordinates are given with respect to ($\alpha_0$, $\delta_0$)$=$($51^\circ$, $-69^\circ$).}
\label{map}
\end{figure*}

The spatial distribution of the $360$ spectroscopically confirmed QSOs behind the Magellanic system is shown in Fig. \ref{map}. Most of them, confirmed prior to July 2009, are included in the compilation by V\'{e}ron-Cetty \& V\'{e}ron (\cite{ver10}) and we use their coordinates, corrected according to Flesch (\cite{fle12}), instead of those from the original studies. The lists of QSOs behind the LMC and SMC are discussed in Sects. \ref{lmcqso} and \ref{smcqso}, respectively. For the presently studied VMC dataset only $117$ quasars are included within the regions observed by the VMC survey, with most of them contained in tiles LMC $5\_5$, LMC $6\_4$, SMC $4\_3$ and SMC $5\_4$. Tile numbering begins from the bottom right corner, increasing from right to left and from bottom to top. The first LMC tile is $2\_3$, the first SMC tile is $2\_2$, the first Bridge tile is $1\_2$ and Stream tile $1\_1$ is right above the Bridge while $2\_1$ is to the right of the SMC. The central coordinates of VMC tiles are given in Paper I while the contour plots showing the outer structure and bar of the galaxies as well as the distribution of evolved giant stars are described in Cioni, Habing \& Israel (\cite{cio00}). 

Figure \ref{map} shows that the distribution of known QSOs is inhomogeneous and biased to the central region of the galaxies. This is because most QSOs were extracted from micro-lensing surveys that have focused their observations on the densest regions, i.e. the bar of the galaxies. In addition, not all candidates from these searches have been spectroscopically followed-up, further limiting the area with known QSOs.

\subsection{Cross-correlation}
\label{match}
VMC detections of the $126$ known (spectroscopically confirmed) QSOs covered by VMC data up to September 2011 were selected by searching for the nearest counterpart within $2^{\prime\prime}$. This produced $117$ matches in the merged catalogue ("vmcsource") containing $Y$, $J$ and $K_\mathrm{s}$ band sources extracted from "tiledeepstacks" which are deep tile images resulting from the combination of individual tile images taken at different observing epochs. A further $8$ QSOs were matched in the un-merged catalogue, and only one QSO was undetected. We inspected the images of all $126$ objects and most were good and well matched to counterparts at $<1^{\prime\prime}$, see Sect. \ref{vmcqsodis} for details.

Table \ref{qsotable} (electronically available)  lists the VMC data for the $117$ known quasars (three are shown as an example). The first column gives the QSOs name from the literature for which references are in column $2$. Columns $3$ and $4$ list the J2000 coordinates of the VMC counterparts. The $Y$, $J$ and $K_\mathrm{s}$ magnitudes and $1$ sigma uncertainties are listed in columns $5-10$. Column $11$ lists the VDFS source classification flag ($1$ for galaxies, $-3$ for probable galaxies, $-1$ for stars and $-2$ for probable stars), column $12$ the VMC tiles where QSOs were found and column $13$ the Flag (see Sect. \ref{data1}). 

Of the $117$ matched QSOs, $67$ lie behind the LMC, $47$ behind the SMC and $3$ behind the Bridge regions. The majority of the QSOs are detected in all three wave bands, with $1$ QSO present only in $YJ$ data, $3$ only in $JK_\mathrm{s}$ data and $2$ only in $K_\mathrm{s}$ data. The VDFS pipeline classifies $\sim53$\% of the matched QSOs as galaxies and the other $47$\% as stars or probable stars. 

For the $8$ QSOs detected in the un-merged VMC catalogue, but blended with other sources which appear in the VMC merged catalogue, Table \ref{qsotableprob} lists the name, reference, wave bands and VMC tile in columns $1$, $2$, $3$ and $4$, respectively. These objects are not used in our analysis.

\begin{table*}
\caption{VMC quasar parameters.}
\label{qsotable}
\[
\begin{array}{lcccccccccrcc}
\hline
\noalign{\smallskip}
\mathrm{Name} & \mathrm{Ref.} & \alpha & \delta & Y & \sigma_Y & J & \sigma_J & K_\mathrm{s} & \sigma_{K_\mathrm{s}} & \mathrm{Class} & \mathrm{Tile} & \mathrm{Flag} \\
 & & \mathrm{(h:m:s)}  & \mathrm{(d:m:s)} & \mathrm{(mag)} & \mathrm{(mag)} & \mathrm{(mag)} & \mathrm{(mag)} & \mathrm{(mag)} & \mathrm{(mag)} & & & \\
 \hline
 \noalign{\smallskip}
 \mathrm{OGLE}\,003850.79$-$731053.1 & 3 & 00$:$38$:$50.80 & -73$:$10$:$53.4 & 16.562 & 0.006 & 16.142 & 0.006 & 14.878 & 0.006 & -1 & \mathrm{SMC}\,4\_3 & \mathrm{A} \\
 \mathrm{J}003857.50$-$741000.7            & 10 & 00$:$38$:$57.53 & -74$:$10$:$00.9 & 17.371 & 0.010 & 17.104 & 0.010 & 16.449 & 0.0123 & -1 & \mathrm{SMC}\,3\_3 & \mathrm{A} \\
 \mathrm{J}003957.65$-$730603.6            & 7 & 00$:$39$:$57.64 & -73$:$06$:$03.6 & 18.321 & 0.022 & 18.334 & 0.026 & 16.900 & 0.019 & 1 & \mathrm{SMC}\,4\_3 & \mathrm{A} \\
 \noalign{\smallskip}
\hline
\end{array}
\]
References. (1) Dobrzycki et al. (\cite{dob02}); (2) Geha et al. (\cite{geh03}); (3) Dobrzycki et al. (\cite{dob03a}); (4) Dobrzycki et al. (\cite{dob03b}); (5) Dobrzycki et al. (\cite{dob05}); (6) V\'{e}ron-Cetty \& V\'{e}ron (\cite{ver10}); (7) Koz{\l}owski et al. (\cite{koz11}); (8) Koz{\l}owski et al. (\cite{koz12}); (9) Tinney et al. (\cite{tin97}); (10) Kamath et al. (in prep.).
\end{table*}

\begin{table}
\caption{VMC blended quasars.}
\label{qsotableprob}
\[
\begin{array}{lccc}
\hline
\noalign{\smallskip}
\mathrm{Name} & \mathrm{Ref.} & \mathrm{Filter} & \mathrm{Tile} \\
 & & & \\
 \hline
 \noalign{\smallskip}
\mathrm{J}004753.62$-$724350.6           & 7 & K_\mathrm{s} &  \mathrm{SMC}\,4\_3 \\
\mathrm{J}005444.70$-$724813.7           & 7 &  K_\mathrm{s} &\mathrm{SMC}\,4\_3 \\
\mathrm{J}005714.13$-$723342.8           & 7 & K_\mathrm{s} & \mathrm{SMC}\,5\_4 \\
\mathrm{MQS\,J}051944.39$-$701957.3 & 8 & J, K_\mathrm{s} & \mathrm{LMC}\,5\_5 \\
\mathrm{MQS\,J}052300.14$-$701831.7 & 8 & Y, K_\mathrm{s} & \mathrm{LMC}\,5\_5 \\
\mathrm{MQS\,J}052431.88$-$702231.6 & 8 & K_\mathrm{s} & \mathrm{LMC}\,5\_5 \\
\mathrm{MQS\,J}052908.79$-$702445.7 & 8 & J, K_\mathrm{s} & \mathrm{LMC}\,5\_5 \\
\mathrm{MQS\,J}053304.31$-$714848.4 & 8 & K_\mathrm{s} & \mathrm{LMC}\,4\_6 \\
\noalign{\smallskip}
\hline
\end{array}
\]
References as in table \ref{qsotable}.
\end{table}

\section{Results}
\label{results}

\subsection{$YJK_\mathrm{s}$ colour-colour diagram and known QSOs}
\begin{figure}
\resizebox{\hsize}{!}{\includegraphics{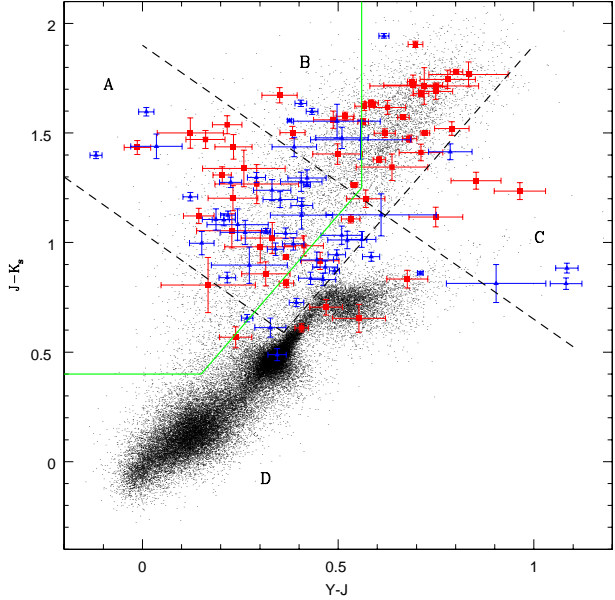}}
\caption{Colour-colour diagram showing the known QSOs classified as galaxy-like (red filled squares) and as star-like (blue filled triangles) in the VMC data. Sources from the LMC tile $8\_8$ with photometric errors $<0.1$ mag and quality flags $=0$ in each wave band are shown with black dots. Dashed black lines identify the regions where known QSOs are found while the green line encloses the region with PNe. 
}
\label{ant}
\end{figure}

Figure \ref{ant} shows the distribution of the $117$ known QSOs in the $YJK_\mathrm{s}$ colour-colour diagram superimposed to the distribution of sources detected in the LMC tile $8\_8$. This tile was chosen for comparison because it is located in the disc of the LMC, where crowding influences the detection of sources rendering the sample with small photometric errors shallower than in the outer regions. In addition, the VMC data for this tile are complete. All $117$ known QSOs are shown regardless of their quality flag and we have checked that the analysis of this paper does not change if only QSOs with Flag A (Sect. \ref{data1}) are used.
Most QSOs are located within the triangle limited by the following lines and the plotted region:
\begin{equation}
(J-K_\mathrm{s}) = -1.25\times(Y-J) + 1.05
\label{one}
\end{equation}
\begin{equation}
\mathrm{and\,} (J-K_\mathrm{s}) = 2.05\times(Y-J) - 0.15 
\label{two}
\end{equation}
\begin{equation}
\mathrm{where\,the\,line\,} (J-K_\mathrm{s}) = -1.25\times(Y-J) + 1.90
\label{three}
\end{equation}
marks the division between star-like (region A) and galaxy-like (region B) QSOs. This line is parallel to the first line (Eq. \ref{one}) that defines a blue colour boundary beyond which no known QSOs are present. The other line (Eq. \ref{two}) runs parallel to the locus of stars and divides them from extra-galactic sources. Some known QSOs have colours outside the  triangular area which is divided into regions C and D by prolonging the line defined by Eq. \ref{three}.
The selection criteria that embrace the sample of known QSOs are not influenced by the choice of tile for the foreground sources, which are only shown for the purpose of visualisation or by reddening (Sect. \ref{reddening}). 

Among other objects populating the colour-colour space, the highest concentrations of non-QSOs are attributed to main-sequence stars at the bluest colours and to giant stars of the LMC. The plume of sources with $(Y-J)>0.5$ mag belongs to the MW. Source at $(Y-J)\sim0.6$ mag and $(J-K_\mathrm{s})\sim1.5$ mag are mostly background galaxies. 
Of the total number of QSOs occupying region A, $65$\% are classified as star-like in VMC data and $35$\% as galaxy-like. QSOs in region B have the opposite distribution, with $21$\% classified as star-like and $79$\% as galaxy like. Many of the star-like QSOs are redder than the green sloped line delimiting the region where most planetary nebulae (PNe) are found (Miszalski et al, \cite{mis11b}, hereafter Paper II). Only a handful of QSOs occupy region C, but  the number of sources in this region does increase in the presence of interstellar reddening, e.g. in tile LMC $6\_6$ (Table \ref{candidates}). The number of QSOs located in region D, that cannot be distinguished from LMC and MW stars by their colours, is small. QSOs in regions C and D are half classified as star-like and half as galaxy-like.

The $YJK_\mathrm{s}$ colour-colour diagram (Fig. \ref{ant}) represents the best VMC diagram to distinguish the bulk of background QSOs from the foreground stellar population of the MCs and the MW. It also provides a clear indication of the location of normal galaxies. We note that our colour selection is analogous to the initial criterion in the KX QSO selection method by Maddox et al. (\cite{mad12}) which uses $gJK$, but with the $g$ band replaced by the $Y$ band. For a similar distribution in the $VJK$ colour-colour diagram see Warren,  Hewett \& Foltz (\cite{war00}).

\subsection{$YJK_\mathrm{s}$ colour-colour diagram and QSO models}

\begin{figure}
\resizebox{\hsize}{!}{\includegraphics{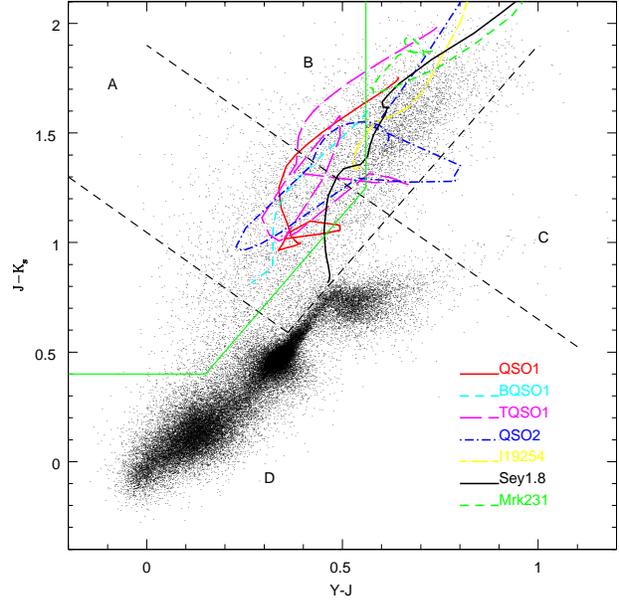}}
\caption{Colour-colour diagram showing models from the SWIRE template library (Polletta et al. \cite{pol07}) and sources as in Fig. \ref{ant}.}
\label{teoant}
\end{figure}

QSO templates obtained from Spitzer-space-telescope Wide-field InfraRed Extragalactic (SWIRE) project template library by Polletta et al. (\cite{pol07}) have been convolved with the VISTA spectral response over each of the $YJK_\mathrm{s}$ filters and the location of these QSOs is shown in Fig. \ref{teoant}. 
The template models are semi-empirical, i.e. they have been constructed combining observational data for similar objects. Those used here refer to three type $1$ QSOs, one type $2$ QSO and two obscured QSOs as well as a moderately luminous AGN. Regions A$+$B where most known QSOs are found are well matched by the models.
The average redshift from the QSO models (excluding I19254 and Sey 1.8) is $1.22\pm0.25$ in region A and $0.44\pm0.25$ in region B. This is consistent with the trends presented in Sect. \ref{redshift}.

\subsection{Colour-magnitude diagrams}

\begin{figure}
\resizebox{\hsize}{!}{\includegraphics{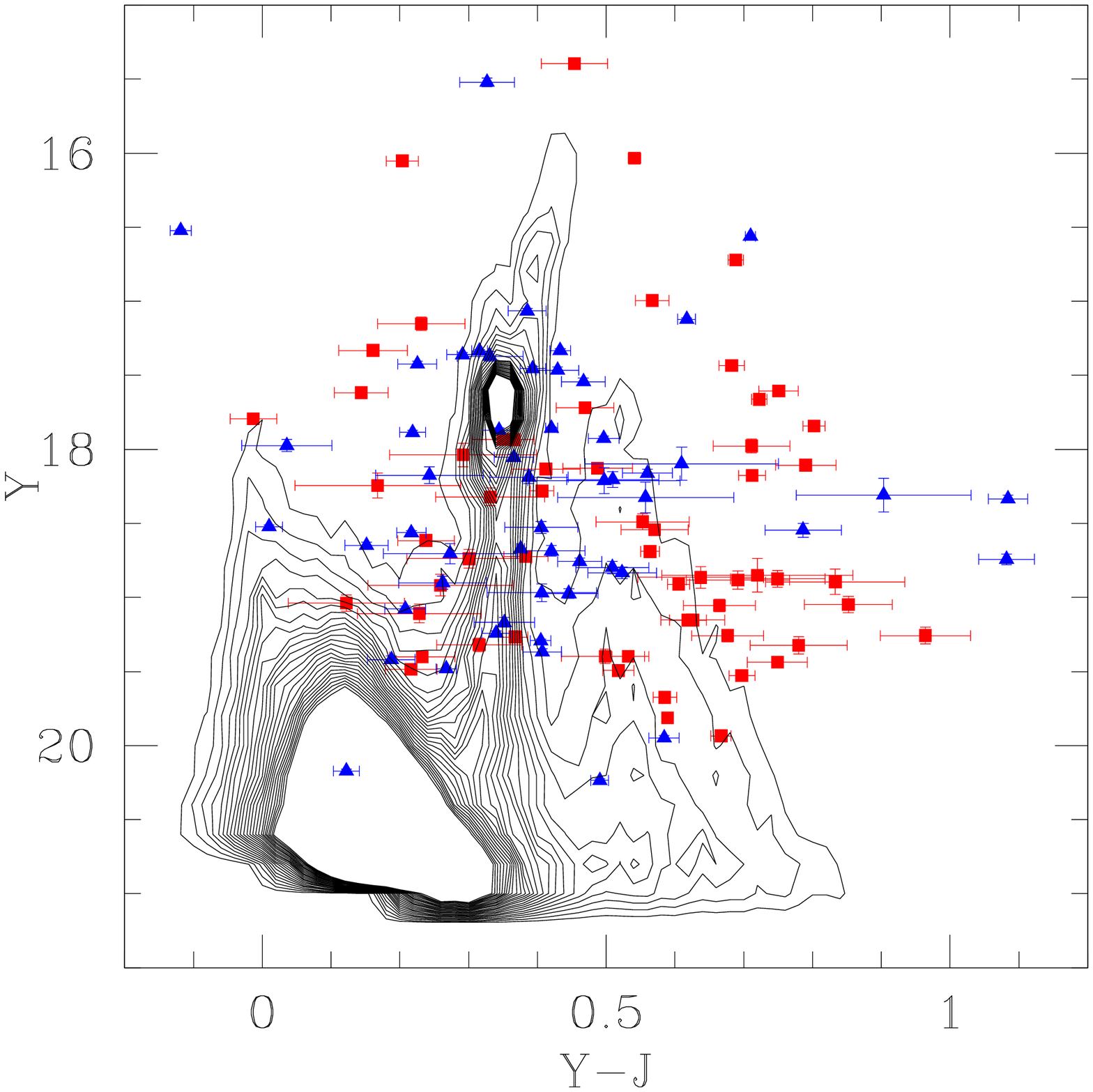}
\includegraphics{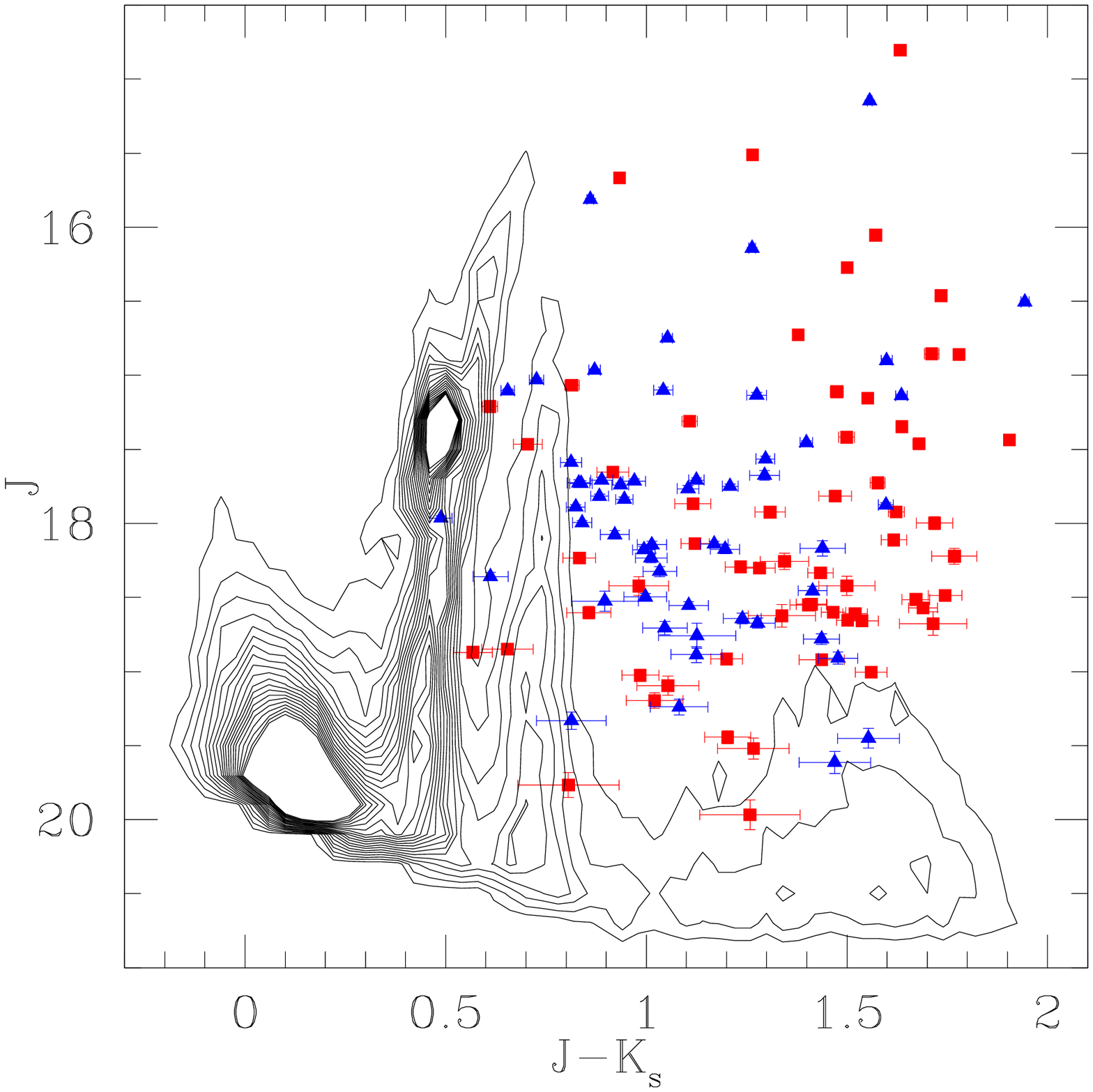}
\includegraphics{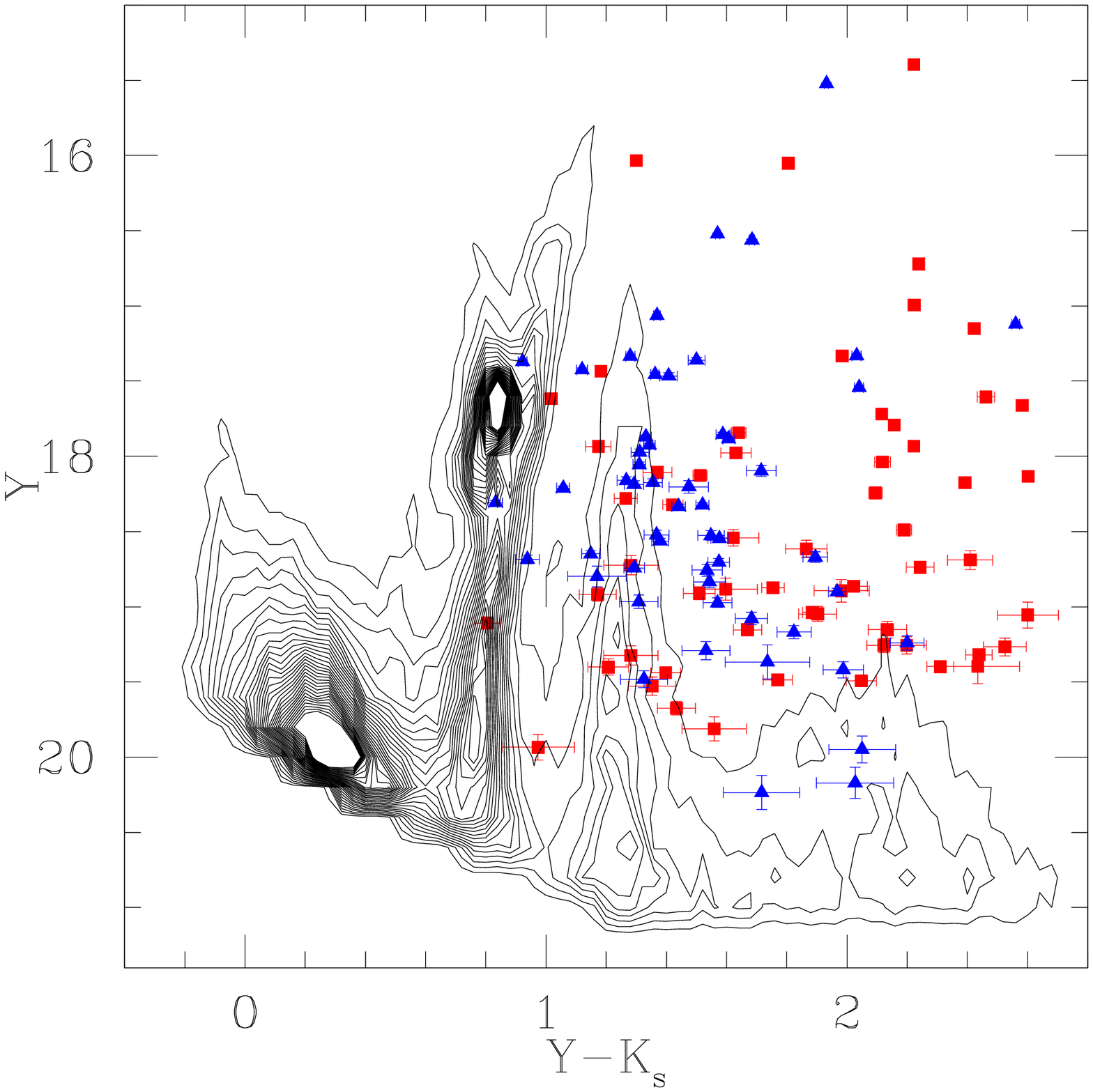}}
\resizebox{\hsize}{!}{\includegraphics{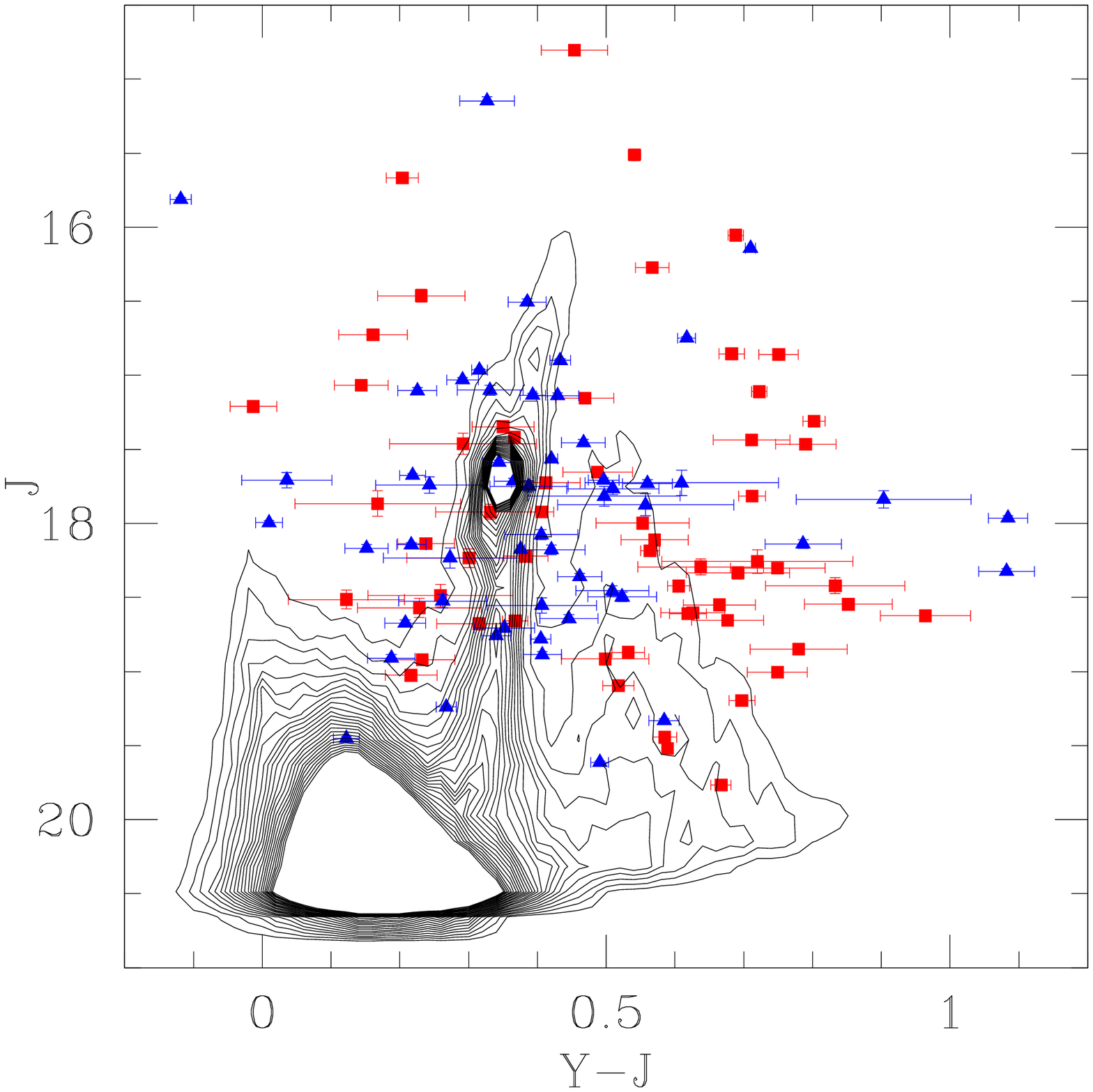}
\includegraphics{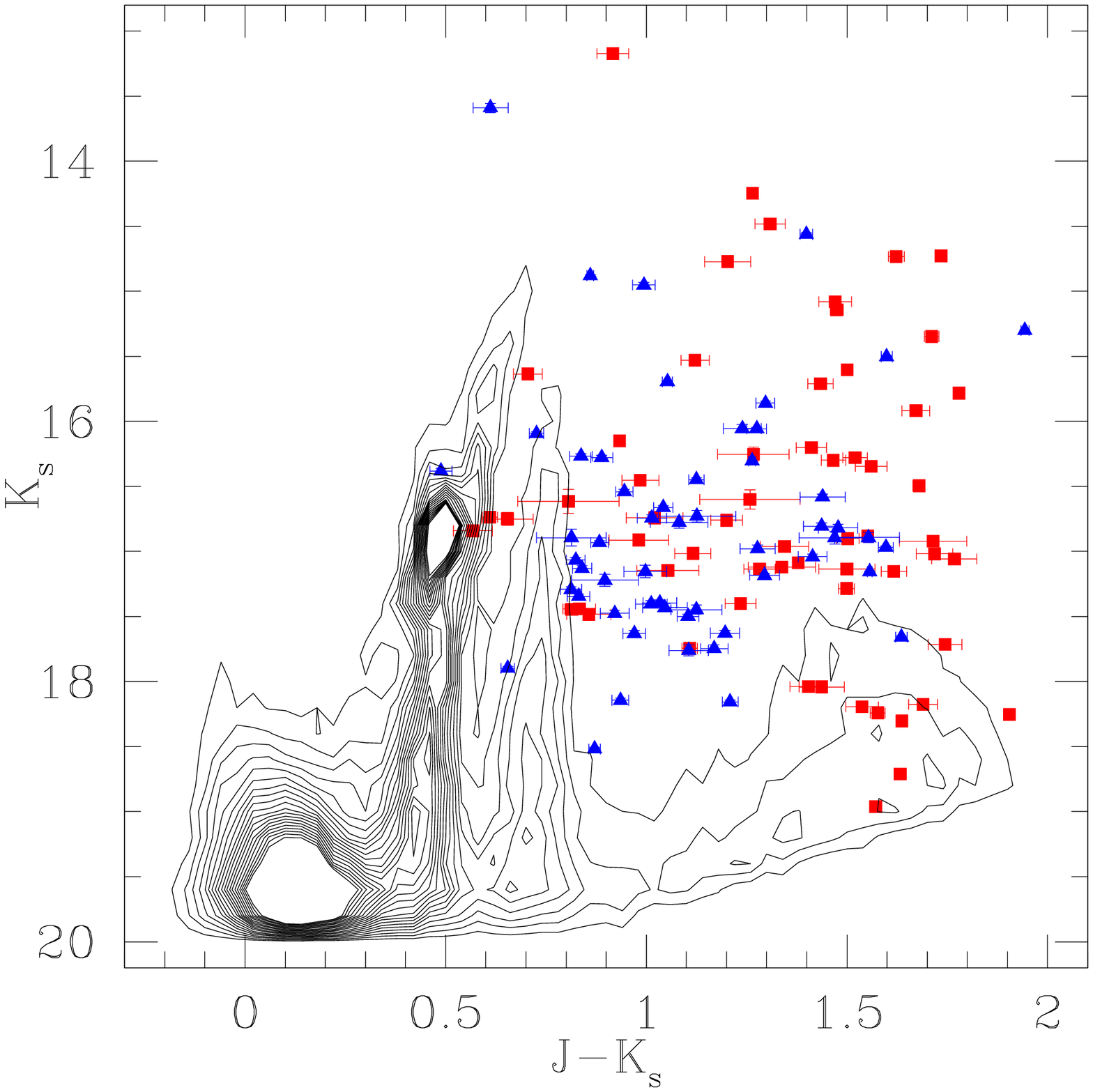}
\includegraphics{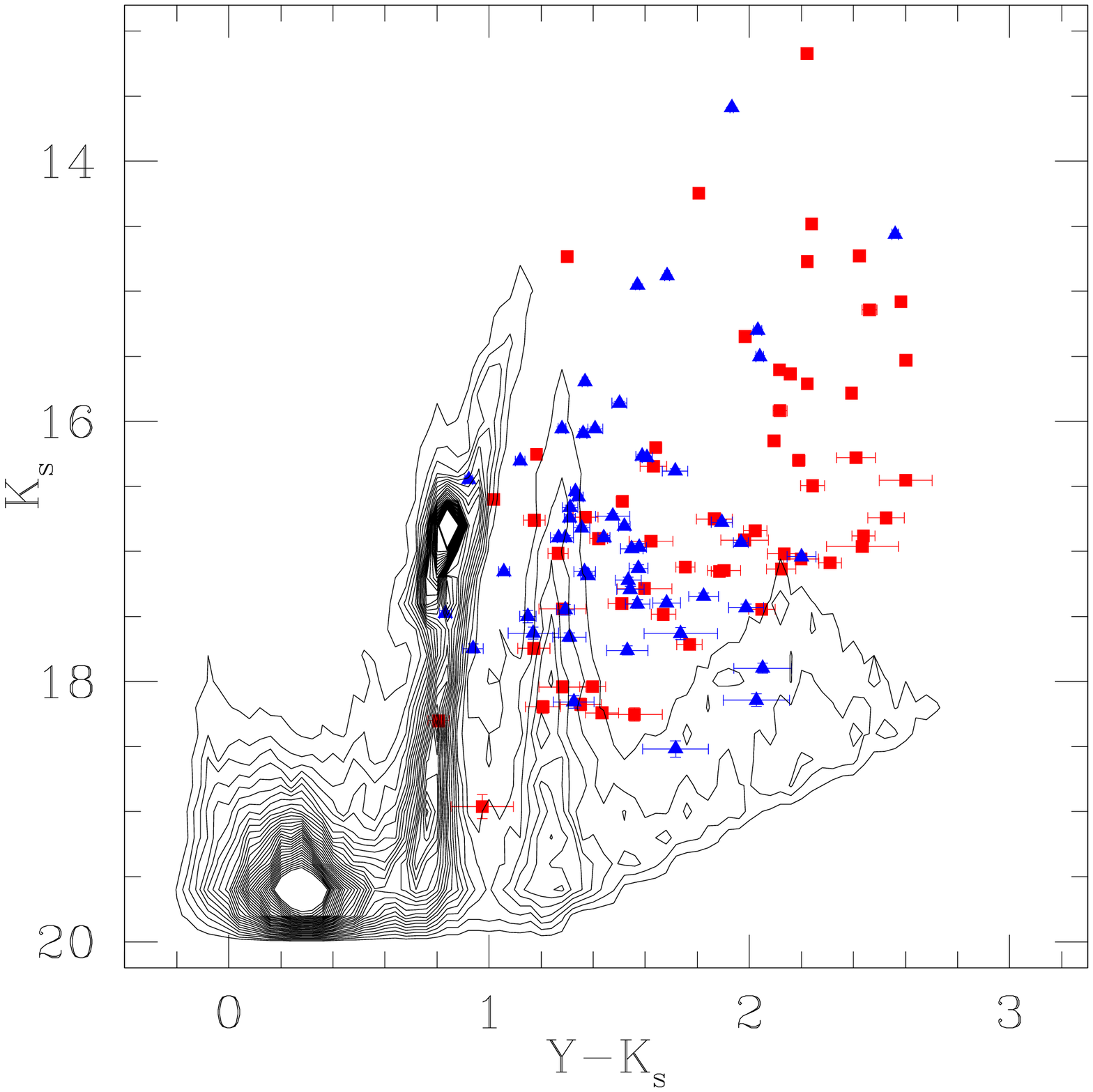}}
\caption{Hess diagrams of sources in tile LMC $8\_8$ with photometric errors $<0.1$ mag and quality flags $=0$, in each pair of two wave bands, shown as contour plots. The known QSOs behind the Magellanic system are displayed as squares. QSOs are colour coded as in Fig. \ref{ant}.}
\label{cmd}
\end{figure}

The distribution of known QSOs in the colour-magnitude diagrams (CMDs) is shown in Fig. \ref{cmd} superimposed on the distribution of LMC and MW sources present in tile LMC $8\_8$. 

The $(Y-J)$ vs. $Y$ or $J$ CMDs show that QSOs overlap with the stellar populations of the LMC and MW. 
The $(Y-K_\mathrm{s})$ vs. $Y$ or $K_\mathrm{s}$ CMDs provide the best separation between the LMC and MW stars as well as background galaxies. QSOs, especially those with a stellar morphology, overlap with MW stars. 
The $(J-K_\mathrm{s})$ vs. $J$ or $K_\mathrm{s}$ CMDs offer instead the best separation between QSOs and stars of the MW or LMC. In the $(J-K_\mathrm{s})$ vs.  $K_\mathrm{s}$ CMD, galaxy-like QSOs overlap with the cone of galaxies (with base $1<(J-K_\mathrm{s})<2$ mag and vertex at $(J-K_\mathrm{s})\sim1.5$ mag); see Kerber et al. (\cite{ker09}), while star-like QSOs seem well separated from it.

Known QSOs in the CMDs unsurprisingly indicate that the present sample is magnitude limited and that the VMC data potentially finds candidates fainter  by $>1$ mag.

\subsection{Variability from $K_\mathrm{s}$ photometry}
\label{secvar}

The VMC survey provides at least $12$ epochs in $K_\mathrm{s}$ observed under similar observing conditions (Paper I). Additional epochs may, however, be obtained in worse conditions. 
The number of epochs currently available (Table \ref{vmcdata}) is not sufficient to sample in detail a QSO light-curve but the accuracy of the photometry is enough to detect significant variability.

\begin{figure}
\resizebox{9cm}{5cm}{\includegraphics{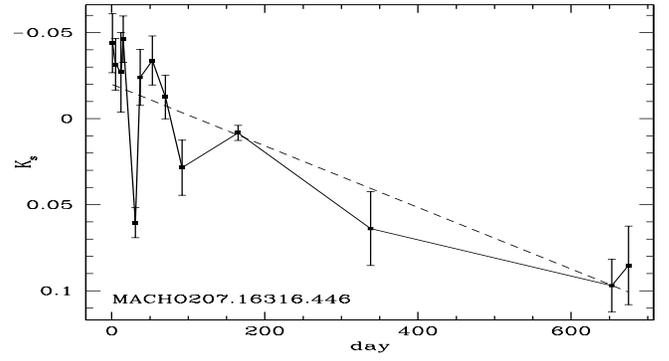}}
\caption{Mean $K_\mathrm{s}$ band QSO variation. Nightly averaged points are connected by a continuous line while a dashed line represents a linear fit through them. Error bars are the standard deviations from the mean. The slope of the line is $1.8\pm0.4\times 10^{-4}$ mag$/$day where day $0$ corresponds to the first observation in the VMC data.}
\label{curve}
\end{figure}

The light-curves for each known QSO from the presently available VMC data are shown in Sect. \ref{qsocurves}, Fig. \ref{curve} shows one example. Good measurements obtained during the same night have been averaged and the error bar corresponds to the standard deviation of the mean. We refer to the $39$\%, $27$\% and $33$\% of the known QSO light-curves sampled over $300-600$ days, $40-80$ days and shorter days as the long, mid and short-range samples, respectively. The current short sample includes some single epoch observations. 

\begin{figure}
\resizebox{9cm}{5cm}{\includegraphics{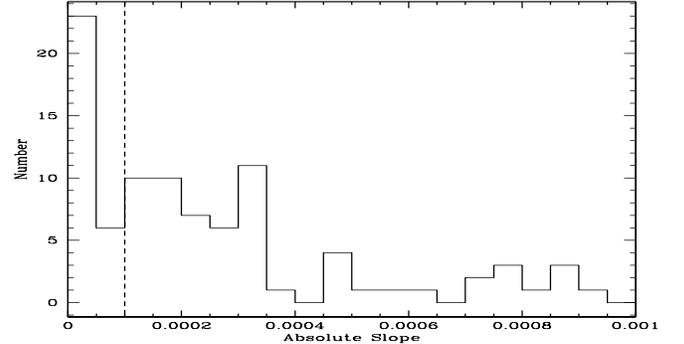}}
\caption{Number of QSOs as a function of slope of variation in the $K_\mathrm{s}$ band. Each bin is $0.5\times10^{-4}$ mag$/$day. A vertical dashed line is drawn at $10^{-4}$ mag$/$day; $75$\% of the QSOs have slope larger than this value.}
\label{slope}
\end{figure}

The shapes of the light-curves are in general irregular without any clear periodicity. There is often a change in  brightness as a function of time both in the long- and mid-range samples. Peaks superimposed on this trend are in some cases significant with respect to the error bars. On average there are broad (up to $100$ days wide) and narrow ($20-40$ days wide) peaks. Curves of the mid-range sample appear noisier but this is largely an effect of the more frequent sampling compared to the long-range sample. The  slope of the overall K$_\mathrm{s}$ variation has been derived for each known QSO and the number distribution of the slope values is shown in Fig. \ref{slope}.  There is a narrow peak due to sources with slope $\sim 0$ mag/day while most of the sample has $1<$ slope $<3.5\times10^{-4}$ mag/day. Only with a larger sample of QSOs we will be able to establish if the latter is consistent with a decreasing gaussian function or represents a bump in the distribution. The remaining QSOs define a tail out to slope $\sim 10^{-3}$ mag/day. In summary $\sim75$\% of the presently known QSOs have slope in the $K_\mathrm{s}$ band $>10^{-4}$ mag/day.

\subsection{Expected number of QSOs}
\label{density}

 In a given $1.5$ deg$^2$ VMC tile we expect $17$ QSOs with $i<19$ mag (Schneider et al. \cite{sch10}).  
 The approximate $i$ band magnitude corresponding to a given $K_\mathrm{s}$ band magnitude has been calculated assuming that the $\sim2.5$ mag difference between the tip of the red giant branch (TRGB) at $K_\mathrm{s}$ and at $I$ is applicable to other sources and that there is not a major difference between the $i$ and $I$ filters. The TRGB in the LMC occurs at $K_\mathrm{s}=12$ mag and at $I=14.54$ mag (Cioni et al. \cite{cio00tip}). 

Tile LMC $5\_5$ is entirely within the area where the known QSO sample is complete to $I<19.2$ mag (Koz{\l}owski et al. \cite{koz12}) and in this tile the faintest known QSO  (MQS J$051953.65-704622.7$) has $Y=19.32$, $J=19.09$ and $K_\mathrm{s}=18.04$ mag. The numbers of known QSOs, the numbers of objects brighter than these limits and the fractions of known QSOs with respect to these numbers in regions A, B and C, respectively, of tile LMC $5\_5$ are listed in Table \ref{candidates}.
Even with the small number statistics it is clear that galaxy-like QSOs dominate region B and star-like QSOs dominate region A.

By taking the ratio of all sources lying in colour selected regions A \& B (Fig. \ref{ant}) found down to the same magnitude as the faintest one and with the same photometric and quality flags to the number of known QSOs, it appears that $\sim6\%$ of all region A \& B sources in tile LMC $5\_5$ are known QSOs. Those in region C, may be influenced by reddened LMC sources and so their percentage should be considered as an upper limit. Using this information it is possible to extrapolate the number of QSOs VMC may expect to find in the other tiles.

\begin{table*}
\caption{VMC quasar candidates.}
\label{candidates}
\[
\begin{array}{lc|ccc|ccc|ccc}
\hline
\noalign{\smallskip}
\mathrm{Sources} & \mathrm{Tile} & \multicolumn{3}{|c|}{\mathrm{Region\,A}^a} & \multicolumn{3}{|c|}{\mathrm{Region\,B}^a} & \multicolumn{3}{|c}{\mathrm{Region\,C}^a} \\
 & & \mathrm{galaxy-like} & \mathrm{star-like} &  \mathrm{all} & \mathrm{galaxy-like} & \mathrm{star-like} & \mathrm{all} & \mathrm{galaxy-like} & \mathrm{star-like} & \mathrm{all}\\
 \hline
 \noalign{\smallskip}
  \mathrm{Known\,QSOs}              & \mathrm{LMC}\, 5\_5  &     4 &   8 &   12 &      5 & 0 &      5 &   2 &   1 &   3\\
  \mathrm{All\,sources^{c}}           & \mathrm{LMC}\, 5\_5  & 137 & 29 & 166 & 140 & 3 & 143 & 14 & 21 & 35\\
  \mathrm{QSO\,fraction\,(\%)}               &  \mathrm{LMC}\, 5\_5 &  2.9 & 27.6 & 7.2 & 3.6 & 0 & 3.5 & 14.3 & 4.8 & 8.6 \\
  \hline
    \mathrm{All\,sources^{c}}          & \mathrm{LMC}\, 6\_6   &  86 & 28 & 114 & 97 & 7 & 104 & 152^b & 845^b & 997^b\\
   \mathrm{Expected\,QSOs}       &  \mathrm{LMC}\, 6\_6   &    2 &   8 &    10 &   3 & 0  &     3 &   22^b &   40^b &    62^b\\  
   \mathrm{Candidate\,QSOs}     & \mathrm{LMC}\, 6\_6    &        &      &    15 &      &      &  11 &             &             &    49^b\\
\hline 
  \mathrm{All\,sources^{c}}          & \mathrm{LMC}\, 8\_8   & 129 & 55 & 184 & 317 & 1  & 318 & 3 & 12 & 15\\
  \mathrm{Expected\,QSOs}         &  \mathrm{LMC}\, 8\_8  &     4 & 15  &   19 &    11& 0  &   11 & 0 &    1 &   1\\
  \mathrm{Candidate\,QSOs}        & \mathrm{LMC}\, 8\_8   &         &       &   14 &         &     &     7 &     &       &   1\\
\hline
\end{array}
\]
$^a$ Regions ABC are those in Fig. \ref{ant}.

$^b$ Includes reddened LMC stars.

$^c$ Objects with $Y<19.32$ mag, $J<19.09$ mag and $K_\mathrm{s}<18.04$ mag, photometric errors $<0.1$ mag and quality flags $=0$ in each band.

\end{table*}

Using Table \ref{candidates}, we now analyse the other two LMC tiles for which VMC observations are completed. Tile LMC $6\_6$ includes the 30 Doradus regions where the reddening is the highest within the LMC implying the presence of many reddened LMC sources in region C. Tile LMC $6\_6$ contains more sources than tile LMC $8\_8$, on average, because of it is nearer to the centre of the galaxy, but the difficulty in detecting them clearly results in larger photometric errors and quality flags. This explains the inferior number of sources selected to have photometric errors $<0.1$ mag and quality flags $=0$ in each wave band (Table \ref{candidates}). Taking all objects in colour regions A \& B and assuming the same percentage of known QSOs, as derived from tile LMC $5\_5$, we find the number of QSOs VMC might expect to detect, irrespective of their morphology, is $13$ in tile LMC $6\_6$ and $30$ in tile LMC $8\_8$ (Table \ref{candidates}).

\subsection{QSO candidates}
\label{vars}

%We have shown that most known QSOs lie in regions A$+$B of the $YJK_\mathrm{s}$ colour-colour diagram. Unfortunately, only $6$\% of all objects in the same region of colour space will actually be QSOs. However we can use the $K_\mathrm{s}$ variability  to select a much tighter sample of candidate QSOs, a much greater fraction of which would be expected to be real QSOs.

As $75$\% of the known QSOs show a variation in the $K_\mathrm{s}$ band with a slope $>10^{-4}$ mag/day (Fig. \ref{slope}) we expect $10$ and $23$ QSOs with this characteristic in tile LMC $6\_6$ and $8\_8$, respectively, among the magnitude, error and quality flag limited samples of candidates. These numbers have been confirmed by examining the light-curve variation of sources in these two tiles, for which all $K_\mathrm{s}$ epochs have been obtained by the VMC survey over a time range of $300-600$ days.

To select objects with variability and colours like the known QSOs, we use the following criteria. Photometric errors $<0.1$ mag and  quality flags $=0$ in $YJK_\mathrm{s}$, a $K_\mathrm{s}$ slope of variation $>10^{-4}$ mag$/$day, $Y<19.32$ mag, $J<19.09$ mag and $K_\mathrm{s}<18.04$ mag. We find a sample of $22$ objects from tile LMC $8\_8$ of which $14$, $7$ and $1$ are located in region A, B and C respectively (Table \ref{candidates}). This number is similar to the expected number of candidate QSOs derived earlier, i.e. $30$ candidate QSOs are expected of which $75$\% ($\sim23$) show a variation in the $K_\mathrm{s}$ band with a slope $>10^{-4}$ mag/day. The same criteria applied to tile LMC $6\_6$ result in a sample of $26$ objects, $15$ and $11$ in regions A and B respectively, while the large number of objects in region C reflects the contamination by reddened LMC stars (Table \ref{candidates}). This suggests that the combination of $YJK_\mathrm{s}$ colours and $K_\mathrm{s}$ variability selection is very effective in making a candidate QSO selection. The number of candidate QSOs would be larger if all sources of photometric class A are used (i.e. including those with $0<$ quality flag $<16$).

\section{Discussion}
\label{discussion}

\subsection{Non-QSO colours}

Photometric criteria for selecting QSOs often result in the misclassification of sources that show similar colours and magnitudes. The main contaminants in candidate QSO samples are normal galaxies, YSOs, PNe and evolved stars. Brown dwarfs have $(Y-J)>0.8$ mag while QSOs have $(Y-J)<0.8$ mag (Warren et al. \cite{war07}). The cool T-type dwarfs have also $-1.5<(J-K_\mathrm{s})<0.5$ mag (Burningham et al. \cite{bur10}) while L-type dwarfs have $(J-K_\mathrm{s})>0.5$ mag (Knapp et al. \cite{kna04}). Only the latter may be present in QSO regions BC (Fig. \ref{ant}). 

\begin{figure}
\resizebox{\hsize}{!}{\includegraphics{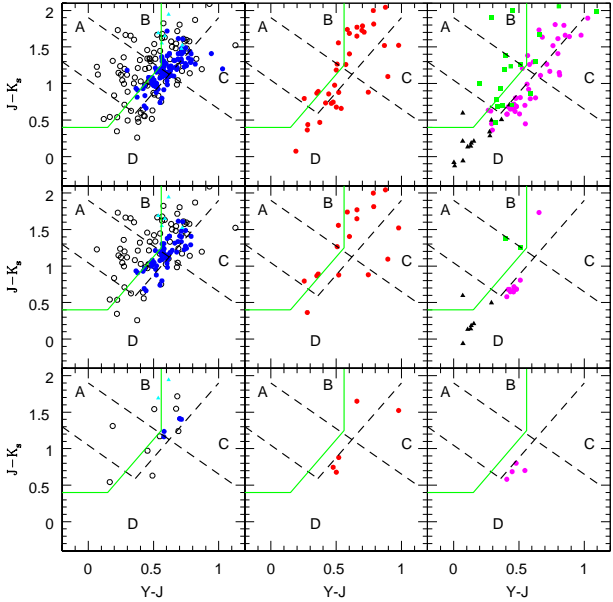}}
\caption{Distribution of the spectroscopically confirmed non-QSOs in the colour-colour diagram from present VMC data as in Fig. \ref{others}.  (left column) Likely QSOs (black circles), normal (blue circles) and Seyfert galaxies (cyan triangles). (middle column) YSOs (red circles). (right column) PNe and post-AGB stars (green squares), red stars (magenta circles) and blue stars (black triangles).}
\label{spectra}
\end{figure}

A list of spectroscopically confirmed non-QSOs was assembled from the investigations by Koz{\l}owski et al. (\cite{koz12}), Woods et al. (\cite{woo11}) and Kamath et al. (in prep.). These objects were  positionally matched to the VMC merged catalogue and counterparts were assigned to the nearest object within $1^{\prime\prime}$. This sample comprises of $106$ galaxies of which $7$ are Seyferts, $41$ YSOs, $9$ PNe, $14$ post-AGB stars, $11$ blue stars (including Be stars) and $37$ red stars (including AGB and red super giant stars). Ninety-four of the likely QSO candidates, with $I>19$ mag and whose spectra are featureless and prevent a secure identification (Koz{\l}owski et al. \cite{koz12}), were also matched with the VMC data. The distribution of these sources in the colur-colour diagram is shown in Fig. \ref{spectra}.

Comparing Fig. \ref{ant} and \ref{spectra}, it seems that QSO candidates selected from the left side of the green line in regions A and B are unlikely to be contaminated by stars, galaxies or YSOs, but there will be a contamination by PNe. QSO candidates selected from region A but to the right of the green line will be contaminated by galaxies and YSOs. QSO candidates selected from region B but to the right of the green line will be contaminated by normal and Seyfert galaxies, YSOs and red stars. The sub-set of variable objects (with slope in the $K_\mathrm{s}$ band $>10^{-4}$ mag/day) weeds out contaminating red stars from regions B and C, and PNe from regions A and B while the distribution of galaxies and YSOs is unchanged. The distribution of objects with photometric errors $<0.1$ mag and quality flags $= 0$ results in a negligible number of galaxies and YSOs in regions B and C while none remain in region A. The Koz{\l}owski et al.'s candidate QSOs populate the colour-colour diagram similarly to known QSOs (Fig. \ref{ant}) and those that remain after applying variability, photometric uncertainty and quality flag criteria, are high confidence candidates also according to our study. QSO candidates selected from region A are not influenced by QSO contaminants, but in region B five objects would not likely be QSOs. These correspond to $\sim20$\% of the number of QSO candidates.

The influence of non-QSOs as estimated from the analysis of the photometrically selected, statistically significant whilst not entirely reliable, sample by Gruendl \& Chu (\cite{gru09}) is described in Sect. \ref{photo}. Results show a $0$\% and $50$\% contamination by non-QSOs in region A and B, respectively.

\begin{figure}
\resizebox{\hsize}{!}{\includegraphics{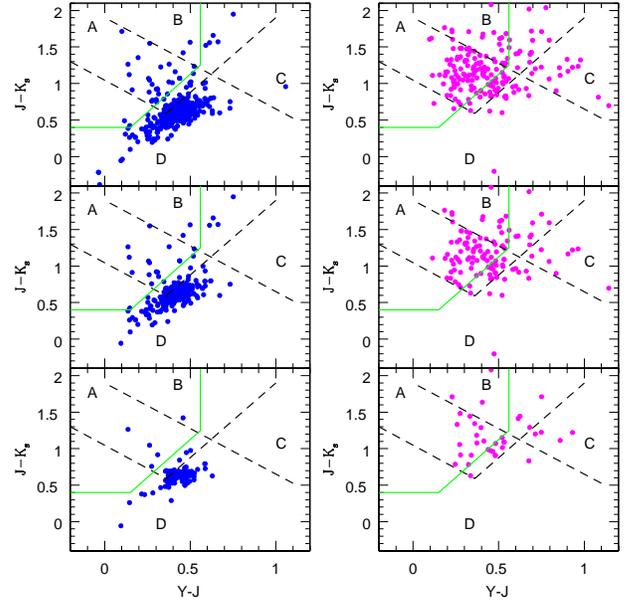}}
\caption{Distribution of low (blue circles) and high (magenta circles) confidence QSOs in the colour-colour diagram from present VMC data as in Fig. \ref{ant}.}
\label{kim}
\end{figure}

A large sample of low and high confidence QSOs was obtained by Kim et al. (\cite{kim12}) as a result of mining the MACHO database. Initially, using a method purely based on the QSO time variability (Kim et al. \cite{kim11}) and later with the additional information from observations at other wavelengths and the development of a model based on the diagnostics of known QSOs (Kim et al. \cite{kim12}). Figure \ref{kim} shows the distribution of the Kim et al.'s candidate samples positionally matched with the VMC data in the colour-colour diagram. There are $574$ low confidence and $179$ high confidence QSOs with a VMC counterpart within $1.5^{\prime\prime}$. The major point to notice is that low confidence QSOs are distributed mainly in region D while high confidence QSOs occupy mostly regions A and B. This is the case in general, for sources with a slope of variation in the $K_\mathrm{s}$ band $>10^{-4}$ mag/day and for sources with photometric errors $<0.1$ mag and quality flags $= 0$ in each wave band strongly supporting the QSO selection criteria developed in this study.

\subsection{QSO variability}

While the optical variability of QSOs is well established (e.g. MacLeod et al. \cite{mac11}) much less is known in the near-IR. The most recent investigation by  Kouzuma \& Yamaoka (\cite{kou12}) addresses the properties of the ensemble variability of QSOs using two data points from the Two Micron All Sky Survey (2MASS; Skrutskie et al. \cite{skr06}) and the DEep Near-Infrared Southern sky survey (DENIS; Epchtein et al. \cite{epc99} ) or the UKIRT Infrared Deep Sky Survey (UKIDSS; Lawrence et al. \cite{law07}). 
About $15-25$\% of the QSO emission in the $K_\mathrm{s}$ band is attributed to the accretion disc (Tomita et al. \cite{tom06}) and the remaining to thermal radiation by dust in the AGN torus, heated by the disc (Barvainis \cite{bav87}). The reason of variations in the $K_\mathrm{s}$ emission may be due to changes in the inner radius of the torus as a consequence of variations in the UV/optical flux from the AGN (Koshida et al. \cite{kos09}, Kawaguchi \& Mori \cite{kaw11}).

The best sampled near-IR QSO light-curves were obtained by the Multicolour Active Galactic Nuclei Monitoring (MAGNUM) programme on NGC 4151 and NGC 5548 (Yoshii, Kobayashi \& Minezaki \cite{yos04}; Koshida et al. \cite{kos09}) or focused on 3C 273 (Courvousier et al. \cite{cor88}). In the first study the $K$ light-curves of the AGN nuclei, after subtracting the contribution from the host galaxy, show clear minima and maxima each over a time range of $\sim 200$ days. Overall the $K$ variation is smoother than the variation in the $V$ band which appears irregular and associated with rapid variations on time scales of several days. The luminous 3C 273 source shows fast and structured flaring events also in the near-IR and only $\sim 15$ days apart. Near-IR variations of QSO emission over a time range of up to twenty years but with a sparse sampling were also found by Neugebauer et al. (\cite{neu89}).

The VMC multi-epoch observations represent a considerable step forward in the study of QSO near-IR variations. The present sample shows both smooth trends, i.e. an increasing and decreasing brightness over a time range of $300-600$ days, and structured events, i.e. broad (up to $100$ days wide) and narrow ($20-40$ days wide) peaks, in agreement with previous studies.  
In the future we plan to construct empirical template light-curves and structure functions (e.g. Hughes, Aller \& Aller \cite{hug92}). These can then be used to refine the search for QSOs and eventually to derive their redshift as in Dai et al. (\cite{dai12}). A theoretical interpretation of the VMC near-IR light-curves and structure functions may also provide further evidence for the mechanism responsible for the variations and contribute to studies of the structure and dynamics of the AGN (e.g. Wold, Brotherton \& Shang \cite{wol07}, Ovcharov et al. \cite{ovc08}).

\subsection{QSO statistics}

Assuming that QSOs are homogeneously distributed across the surface of the Magellanic system we can use the number of VMC candidate QSOs derived in the previous sections to estimate the total number of QSOs detectable by the VMC survey. Using the magnitude limited sample in tile LMC $8\_8$, which is not affected by interstellar reddening and crowding,  we derived that $22$ objects are candidate QSOs of which $\sim 4$ may not be true QSOs. The VMC survey comprises of $110$ tiles of which $68$ in the LMC, $27$ in the SMC, $13$ in the Bridge and $2$ in the Stream areas. The expected QSO population accessible to the VMC survey is thus of at least $1\,200$ objects behind the LMC, $400$ behind the SMC, $200$ behind the Bridge and $30$ behind the Stream areas. This means that $\sim 80$\% of the VMC accessible QSO population behind the Magellanic system is yet to be found. In addition, these numbers are conservative limits since they have originated from the application of magnitude, error, quality flag and variability restrictions. The fraction of QSO candidates suitable for astrometry, individually or as a group, will be explore in a subsequent paper.

\section{Conclusions}
\label{conclusions}

A large sample of QSOs across the Magellanic system is of fundamental importance as an astrometric reference source for investigations of the proper motion, of the interstellar and intergalactic medium along the line of sight and to address the nature of the extra-galactic sources themselves. 

We present criteria for selecting candidate QSOs based solely on the multi-epoch near-IR photometry from the VMC survey developed from the properties of $117$ known QSOs. Most of them occupy two neighbouring regions of the $(Y-J)$ vs. $(J-K_\mathrm{s})$ diagram, one is dominated by QSOs with a star-like appearance and the other by QSOs with a galaxy-like appearance in the VMC data. The analysis of $K_\mathrm{s}$ multi-epochs shows that $\sim75$\% of the QSOs have a slope of variation $>10^{-4}$ mag/day across a time range up to $600$ days. The combination of $YJK_\mathrm{s}$ colours and $K_\mathrm{s}$ variability criteria, with appropriate data quality flags, has been used to define samples of candidate QSOs with a $\sim20$\% contamination by non-QSOs on average, but the contamination is null in the region dominated by objects with a star-like appearance.

In LMC tile $6\_6$, including the 30 Doradus region, and $8\_8$, including the South Ecliptic Pole region, we find $22$ and $26$ QSO candidates with photometric errors $<0.1$ mag and quality flags $=0$ in each VMC wave band, brighter than $19.32$, $19.09$ and $18.04$ mag in the $Y$, $J$ and $K_\mathrm{s}$ band, respectively, and with a $K_\mathrm{s}$ slope of variation $>10^{-4}$ mag/day. The selection of QSOs from the VMC data is in excellent agreement with the sample of high confidence QSOs by Kim et al. (\cite{kim12}).

PNe represent the major source of contamination in the regions of the colour-colour diagram dominated by star-like QSOs while normal galaxies and YSOs are the major contaminants in the region dominated by galaxy-like QSOs. The sub-set of sources with a slope of variation in the $K_\mathrm{s}$ band $>10^{-4}$ mag/day is not influenced by PNe and red stars. The sub-set of sources with photometric errors $<0.1$ mag and quality flags $= 0$ in each VMC wave band is very small. Since known QSOs may be de-blended and have quality flags $=16$ the latter represents a reliable whilst not complete sample of candidates.

VMC magnitudes and $K_\mathrm{s}$ light-curves for known QSOs are provided in a table and in the appendix. The full sample of known QSOs comprises $332$ QSOs from the literature and $28$ newly confirmed QSOs by Kamath et al. (in prep.). VMC data on a fully observed tile in the outer LMC disc ($8\_8$) have been used to estimate the number of QSO candidates behind the entire system as covered by the VMC survey. This is of the order of $1\,200$ behind the LMC, $400$ behind the SMC, $200$ behind the Bridge and $30$ behind the Stream areas. The VMC survey is the most sensitive near-IR survey of the Magellanic system to date providing for the first time near-IR counterparts to many QSOs. Spectroscopic observations of the candidates identified here will support an extension of the selection to faint magnitudes exploiting the whole VMC range available.

\begin{acknowledgements}
MRC acknowledges support from the Alexander von Humboldt Foundation. This research has made use of the SIMBAD database operated at CDS, Strasbourg, France. We are grateful to ESO staff for scheduling and making the VMC observations, the Cambridge Astronomy Survey Unit and the Wide Field Astronomy Unit for providing us with the reduced data and catalogues.
We thank M.I. Moretti for giving comments that improved the clarity of the paper.
\end{acknowledgements}

\begin{appendix}
\label{app}

\section{Known QSOs behind the Magellanic system}
\subsection{LMC}
\label{lmcqso}

The list of $233$ spectroscopically confirmed quasars behind the LMC comprises $75$ QSOs from the V\'{e}ron-Cetty \& V\'{e}ron (\cite{ver10}) compilation among which $10$ were identified as X-ray sources with the Chandra X-ray Observatory satellite and $3$ as a result of mining OGLE-II light-curves (Dobrzycki et al. \cite{dob02}, \cite{dob05}) while $38$ refer to variability in the MACHO database (Geha et al. \cite{geh03}). In addition, two QSOs used by Anguita et al. (\cite{ang00}) and one by  Pedreros et al. (\cite{ped02}) to study the proper motion of the LMC are attributed to a private communication by J. Maza in 1989 and/or do not appear in previous studies.
More recently, $1$ QSO was confirmed by Hony et al. (\cite{hon11}) and $145$ QSOs by Koz{\l}owski et al. (\cite{koz12}), the latter identified from OGLE-III light-curves.
 The entire MACHO database, with light-curves spanning a time range of $\sim 7.5$ yr, was searched by Kim et al. (\cite{kim11}, \cite{kim12}) who trained a support vector machine model with diagnostic features  based on mid-IR colours, spectral energy distribution red-shifts and X-ray luminosity of previously known QSOs to identify $663$ high confidence candidates, but none of them are at present spectroscopically confirmed. Finally, as part of a spectroscopic study on post asymptotic giant branch (AGB) stars, Kamath et al. (in prep.) confirmed $6$ QSOs and detected one broad emission line in $3$ additional objects that may also be QSOs.

%Blanco \& Heathcote (\cite{bla86}) found a QSO in their near-IR grism survey aimed at characterising red giant stars. 

\subsection{SMC}
\label{smcqso}
The list of $100$ spectroscopically confirmed quasars behind the SMC comprises $41$ QSOs from the V\'{e}ron-Cetty \& V\'{e}ron (\cite{ver10}) compilation among which $16$ were discovered as counterparts to X-ray sources (Tinney, Da Costa \& Zinnecker \cite{tin97}; Tinney \cite{tin99}; Dobrzycki et al. \cite{dob03b}), $3$ were identified from their OGLE-II light-curves (Eyer \cite{eye02}; Dobrzycki et al. \cite{dob03a}) and $9$ from their MACHO light-curves (Geha et al. \cite{geh03}).
More recently, Koz{\l}owski et al. (\cite{koz11}) inspected mid-IR colours from the Spitzer Space Telescope and the OGLE-II light-curves of sources in the central $1.5$ deg$^2$ of the SMC. They confirmed $29$ QSOs and showed broad emission lines, typical of QSOs, in $12$ additional objects but the signal to noise ratio (S/N) in their spectra was of lower quality.
The newest QSOs are from Kamath et al. (in prep.) where $17$ are confirmed and $1$ is a likely candidate since only one typical QSO broad emission line was detected in its spectrum.

\subsection{VMC-QSO counterparts}
\label{vmcqsodis}
Forty QSOs in the LMC from the list by Koz{\l}owski et al. (\cite{koz12}) have counterparts at a mean separation on the sky of $0.33\pm0.17^{\prime\prime}$. Five matches were rejected because the corresponding VMC source, contrary to the distribution of most counterparts, was found at $>1^{\prime\prime}$ and in all five cases inspection of the VMC images showed another, fainter, object with the typical red colour of QSOs present but not found in the list of nearby objects. A cross-correlation with the "vmcdetection" un-merged catalogue shows that the likely counterparts are indeed located at $\le 1^{\prime\prime}$ (Table \ref{qsotableprob}). The counterparts to Kamath et al.'s sources have a similar separation on the sky, i.e. $0.31\pm0.17^{\prime\prime}$.

Seventeen QSOs were matched with the sample by Dobrzycki et al. (\cite{dob02}, \cite{dob03a}, \cite{dob03b}, \cite{dob05}) of which $7$ in the SMC and $10$ in the LMC. Thirty SMC QSOs were matched from the lists of Koz{\l}owski et al. (\cite{koz11}); $20$ from their confirmed list and $10$ from their plausible list. Three matches were rejected because their detection in $Y$ and $J$ is suspicious; they are faint and close to a brighter object, and their suggested counterparts are at $>1.3^{\prime\prime}$ on average. The likely counterparts to these QSOs were found in the un-merged catalogue (Table \ref{qsotableprob}). 
The VMC counterpart to QSOs J$004818.76$-$732059.6$, J$004831.50$-$732339.9$, J$010057.77$-$722230.8$ and J$010137.52$-$720418.9$ at $\sim 1^{\prime\prime}$ are perhaps affected by the extended nature of the sources since all other counterparts have a smaller separations on the sky.  Four QSOs were matched with the compilation by V\'{e}ron-Cetty \& V\'{e}ron (\cite{ver10}) of which $3$ in the Bridge and $1$ in the LMC; the latter, RXS J$05466-6415$, is the brightest near-IR quasar in the sample with $K_\mathrm{s}=13.590\pm0.004$ mag. These QSOs were originally confirmed by Perlman et al. (\cite{per98}) and Healey et al. (\cite{hea08}) for the Bridge and LMC, respectively. 

The mean separation on the sky of VMC counterparts for quasars from Dobrzycki et al. (\cite{dob02}, \cite{dob03a}, \cite{dob03b}, \cite{dob05}),  Koz{\l}owski et al. (\cite{koz11}) and V\'{e}ron-Cetty \& V\'{e}ron (\cite{ver10}) are very similar and correspond to $0.64\pm 0.30^{\prime\prime}$. The larger separation of VMC counterparts to the quasars from Koz{\l}owski et al. (\cite{koz11}) compared to the quasars from Koz{\l}owski et al. (\cite{koz12}) is most likely due to their identification from the OGLE database part II in one case and part III in the other. OGLE-II data were obtained with a CCD camera with a scale of $0.4^{\prime\prime}$/pixel while OGLE-III data were obtained with a new camera with a scale of $0.26^{\prime\prime}$/pixel and this may introduce a systematic offset in the world-coordinates between the two systems. Both OGLE surveys are well-matched to the VISTA camera which has a scale of $0.34^{\prime\prime}$/pixel.

Finally, from the list of Geha et al. (\cite{geh03}) $16$ QSOs, of which $2$ in the SMC, were successfully matched. MACHO $061.08072.0358$ was excluded because two VMC candidate sources are present, one at $1.36^{\prime\prime}$ and one at $1.49^{\prime\prime}$. The mean separation on the sky of the VMC counterparts is $0.86\pm0.29^{\prime\prime}$.

%With the exception of one QSO, as mentioned above, all known QSOs falling into VMC observed tiles were matched with the merged or un-merged catalogues. In our analysis we consider only those found in the merged catalogue since the magnitudes of those in the un-merged catalogue may be influenced by the presence of bright neighbours.

\subsection{Correction for reddening}
\label{reddening}
The QSOs and stars in the MCs as well as stars in the MW are reddened by interstellar dust present along the line of sight. Reddening values in terms of $E(V-I)$ were extracted from the extinction map from Haschke et al. (\cite{has11}) for all but $10$ (located outside the map) of the $117$ QSOs. Using the Cardelli, Clayton and Mathis (\cite{car89}) extinction law, the average QSO extinction is $A(Y)=0.045$ mag, $A(J)=0.032$ and $A(K_\mathrm{s})=0.013$ mag. The estimated reddening, while accounting for the MW and LMC components, does not account for the intergalactic dust that lies between the LMC and the QSOs and for dust in the torus surrounding the AGN.

In tile LMC $8\_8$, which is outside the region studied by Haschke et al. (\cite{has11}), reddening values were extracted from the Zaritsky et al. (\cite{zar04}) extinction map. Here, instead of de-reddening each individual star, an average value referring to cool stars only, $A(V)=0.35$ mag, was obtained in a region centred on the tile and with a radius of $12^\prime$. Cool stars were chosen because they have similar characteristics to red clump giant and RR Lyrae stars used by Haschke et al. (\cite{has11}). The size of the region, whilst being smaller than the extent of a tile, is representative of the average extinction since this tile is located in the outer disc of the LMC where the reddening is uniform. By applying the extinction law from Cardelli et al. (\cite{car89}) extinction values of $A(Y)=0.14$ mag, $A(J)=0.10$ mag and $A(K_\mathrm{s})=0.04$ mag were derived. These values may be too high since in the less populated regions of the LMC differences in $E(V-I)$ between Haschke et al. (\cite{has11}) and Zaritsky et al. (\cite{zar04}) amount to $0.07-0.10$ mag on average. In addition, while these values are applicable to LMC stars they are certainly too high towards foreground MW stars and too low towards extra-galactic sources.

In view of the uncertainty in the extinction of LMC tile $8\_8$, the Schlegel, Finkbeiner \& Davis (\cite{sch98}) map based on DIRBE/IRAS mid-IR observations was used to find average reddening and extinction, $E(B-V)=0.067$ and $A(V)=0.207$ mag, in a region of $2$ deg$^2$ in size centred on the. The extinction value is consistent with the mean value, $A(V)=0.187\pm0.040$ mag, derived in Paper IV from the recovering of the star formation history. $A(Y)=0.083$ mag, $A(J)=0.058$ mag and $A(K_\mathrm{s})=0.024$ mag are then obtained. These values are $0.02-0.06$ mag smaller than those derived from the Zaritsky et al. (\cite{zar04}) map. Using the TRILEGAL code (Girardi et al. \cite{gir05}) to simulate the MW population we obtain that it accounts for $A(V)=0.05$ of absorption implying that most of the reddening towards the LMC is due to the galaxy itself.

\subsection{Trends with redshift}
\label{redshift}

\begin{figure}
\resizebox{\hsize}{!}{\includegraphics{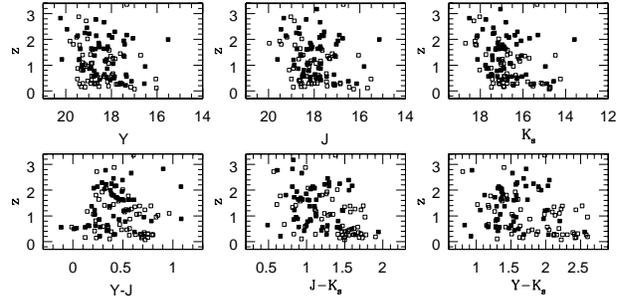}}
\caption{Distribution of redshift versus magnitudes (top line) and colours (bottom line) for known QSOs behind the Magellanic system. Filled symbols indicate sources classified as star-like in the VMC data and empty symbols those classified as galaxy-like.}
\label{red}
\end{figure}

Figure \ref{red} shows the distribution of redshifts with respect to VMC magnitudes and colours. There appears to be no trend as a function of magnitude nor of $(Y-J)$ colour. The QSOs occupy the whole of the parameter space delimited on one side by the bright limit of the survey and on the other by the faint limit of the present sample. The faint limit is also influenced by the incomplete observations of $K_\mathrm{s}$ epochs. 

The plots of redshift as a function of  $(J-K_\mathrm{s})$ and $(Y-K_\mathrm{s})$ colours suggest a trend of larger redshifts at smaller colours. This indicates that QSOs at small redshifts are redder than QSOs at larger redshifts as predicted by Maddox \& Hewett (\cite{mad06}). These authors found QSOs getting bluer as redshift goes from $0$ to $3$ in their simulations.  Figure \ref{red} also shows a dichotomy between star-like and galaxy-like QSOs, concentrated at small redshifts  as in Maddox et al. (\cite{mad08}).

\section{Photometrically selected non-QSOs}
\label{photo}

\begin{figure}
\resizebox{\hsize}{!}{\includegraphics{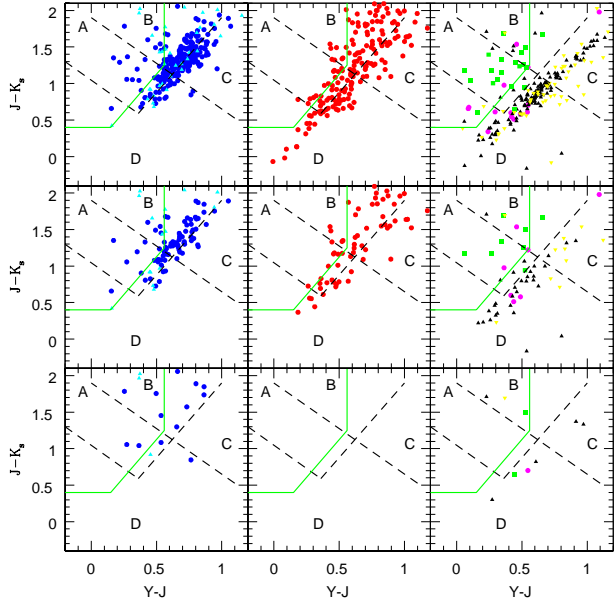}}
\caption{Distribution of the photometrically selected non-QSOs in the colour-colour diagram from present VMC data. (top row) All positionally matched sources. (middle row) Sources with a slope of variation in the $K_\mathrm{s}$ band $>10^{-4}$ mag/day. (bottom row) Sources with photometric errors $<0.1$ mag and quality flags $=0$ in the $Y$, $J$ and $K_\mathrm{s}$ bands. (left column) Definite (blue circles) and probable (cyan triangles) background galaxies. (middle column) YSOs (red circles). (right column) PNe (green squares), diffuse sources (yellow reversed triangles), evolved stars (magenta circles) and other stars (black triangles). Lines are as in Fig. \ref{ant}.}
\label{others}
\end{figure}

A statistically significant sample of QSO contaminants was assembled by Gruendl \& Chu (\cite{gru09}) who analysed spectral energy distributions, from optical to mid-IR wavelengths, and mid-IR images from Spitzer in the central $7^\circ \times7^\circ$ area of the LMC. Their sample has been positionally matched with the present VMC data except for their lists of probable and possible YSOs since it is indicated that the first group may be background galaxies and the second group cannot be definitely classified. 
The distribution of the matched sources in the colour-colour diagram is shown in Fig. \ref{others}. There are: $286$ definite and $48$ probable galaxies, $396$ YSOs, $25$ PNe, $123$ diffuse sources, $44$ evolved stars and $165$ stars. The majority of these sources have counterparts within $\sim1^{\prime\prime}$ compared to an angular resolution of Spitzer of $\sim 2^{\prime\prime}$ (Meixner et al. \cite{mei06}). As expected, galaxies populate region A and B but most of them are located around the line separating them from stars. This is also the preferred location of YSOs which extend to bluer colours and into region B. Most PNe are well confined in the region defined in Paper II while stars and diffuse sources occupy mostly regions C and D. A possible AGB star in the PNe region was found to be a PN by Miszalski et al. (\cite{mis11a}) and its colours are $(Y-J)=0.36$ and $(J-K_\mathrm{s})=2.16$ mag.

The sample of background galaxies by Gruendl \& Chu (\cite{gru09}) contains a mixture of marginally resolved extended sources and extended sources with a point source nuclear region which were obtained from selection criteria tuned to find YSOs. They are therefore not representative of the background galaxy population in general. Among the sample of definite galaxies within the PN region Kamath et al. (in prep.) finds one QSO and one YSO out of three sources in common; the third one could not be classified due to low S/N spectral features. These findings suggest that the Gruendl \& Chu's classification, which is based on photometric criteria, may not be entirely reliable.
Sources with a slope of variation in the $K_\mathrm{s}$ band $>10^{-4}$ mag/day represent a subset of the samples of QSO contaminants and their distribution in the colour-colour diagram is similar to that of the full samples (Fig. \ref{others}).

By considering only sources with photometric errors $<0.1$ mag and quality flags $=0$ in each wave band, all YSOs are excluded (including the one above), only one star, a diffuse source and a PN remain in region B as well as two stars in region C while the number of background galaxies is also considerably reduced and confined within regions A and B, but for one source (Fig. \ref{others}). The main criterion responsible for the removal of sources is the requirement on the quality flags. In fact, the magnitudes of most sources were extracted following a de-blending process corresponding to flag $=16$ (the confirmed QSO above has indeed this quality flag value in each VMC wave band). Since the Gruendl \& Chu's sample is un-confirmed, the remaining galaxies are either QSO contaminants or QSO candidates themselves. These galaxies are not previously known, all but one have a well defined star-like appearance in the VMC images and about half of them have a slope of variation in the $K_\mathrm{s}$ band $>10^{-4}$ mag/day. The latter sample of high confidence QSO candidates, according to our criteria, has a $0$\% and $50$\% influence in region A and B, respectively, of QSO contaminants.

\section{$K_\mathrm{s}$ band light-curves of known QSOs}
\label{qsocurves}

Figure \ref{qsocurves1} shows the $K_\mathrm{s}$ band light-curves of known QSOs in the presently available VMC data (VSA data release VMCv20120126) where single epoch data are included for completeness.

\begin{figure*}
\resizebox{\hsize}{!}{\includegraphics{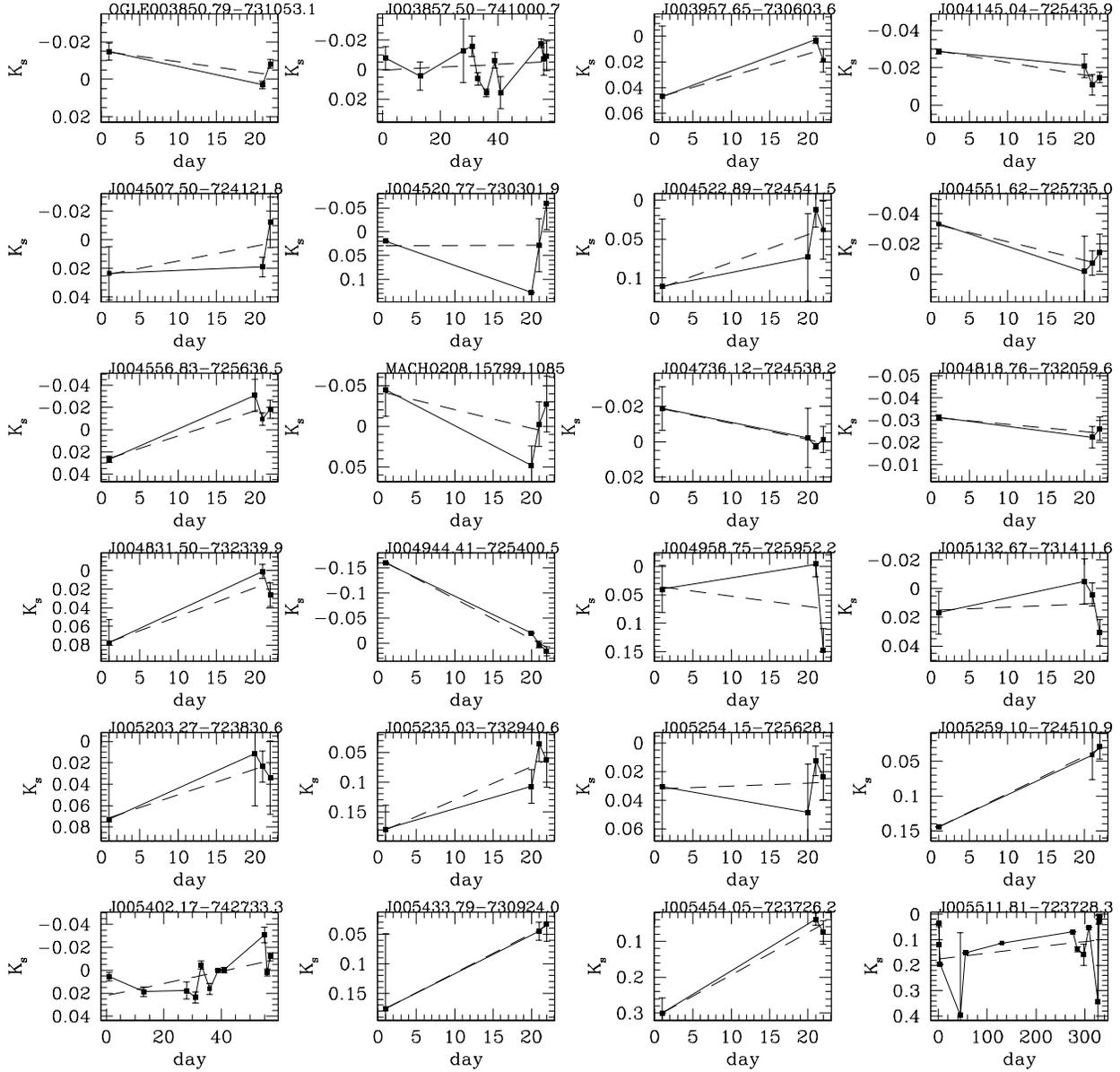}}
\caption{$K_\mathrm{s}$ band light-curves of known QSOs with respect to their mean magnitude. From top to bottom and left to right they follow the order in Table \ref{qsotable}. Points represent the simple average of observations obtained within the same night/epoch and error bars correspond to the error of the mean. A continuous line connects the data points while a dashed line represents a linear fit through them. Day $0$ corresponds to the first observation in the VMC data. Single epoch data are included for completeness.}
\label{qsocurves1}
\end{figure*}

\setcounter{figure}{0}
\begin{figure*}
\resizebox{\hsize}{!}{\includegraphics{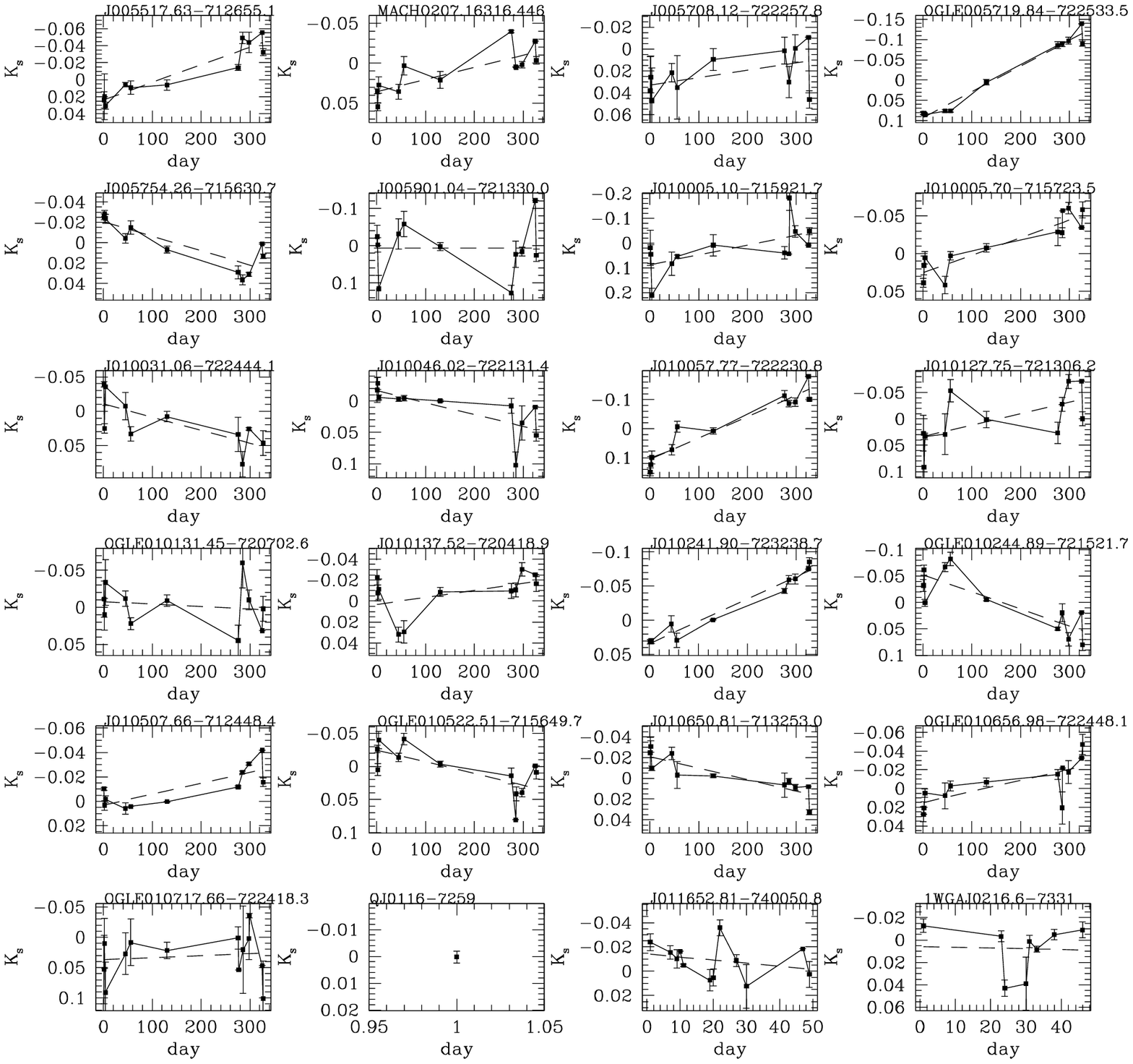}}
\caption{continued.}
\label{qsocurves2}
\end{figure*}

\setcounter{figure}{0}
\begin{figure*}
\resizebox{\hsize}{!}{\includegraphics{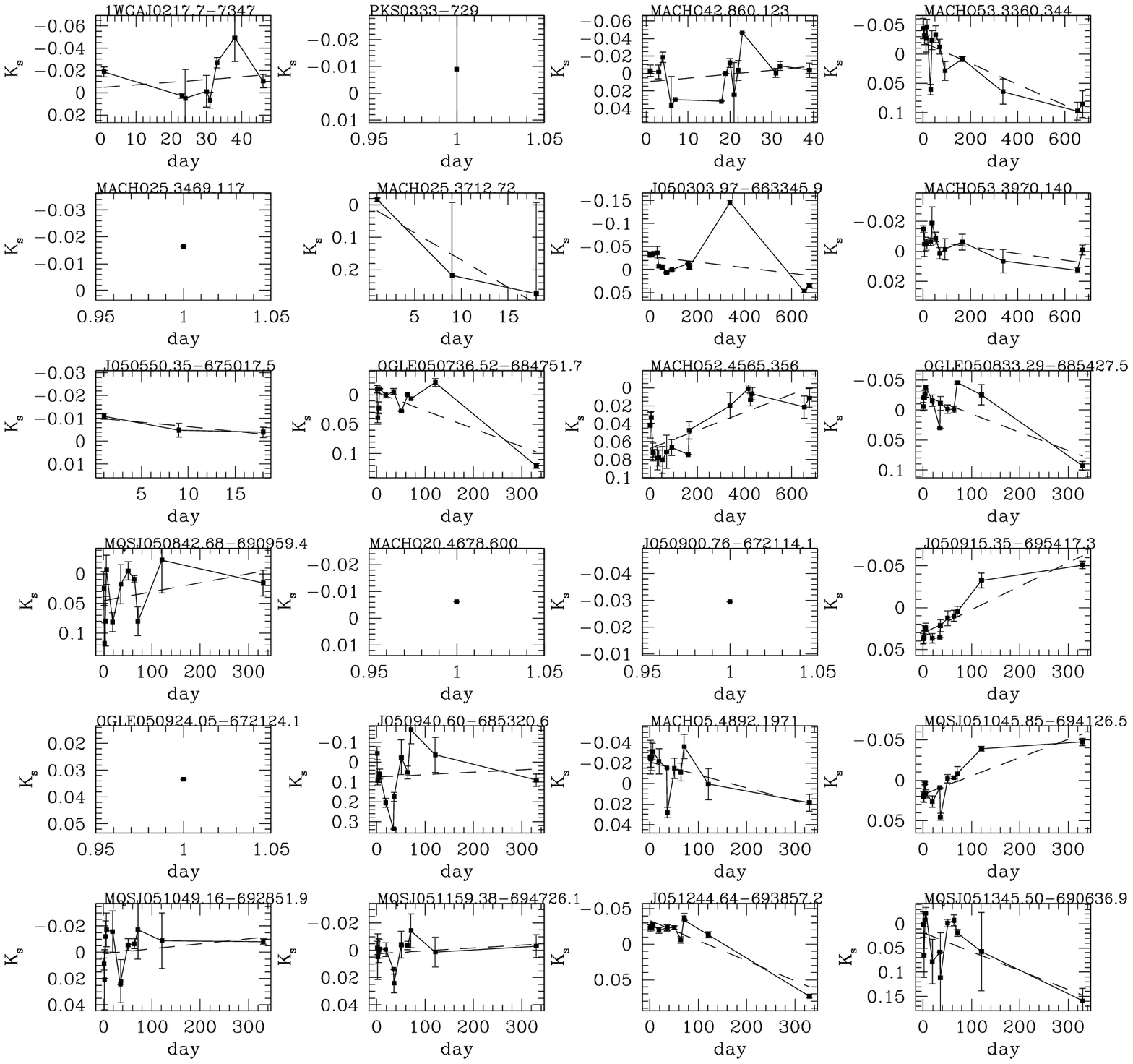}}
\caption{continued.}
\label{qsocurves3}
\end{figure*}

\setcounter{figure}{0}
\begin{figure*}
\resizebox{\hsize}{!}{\includegraphics{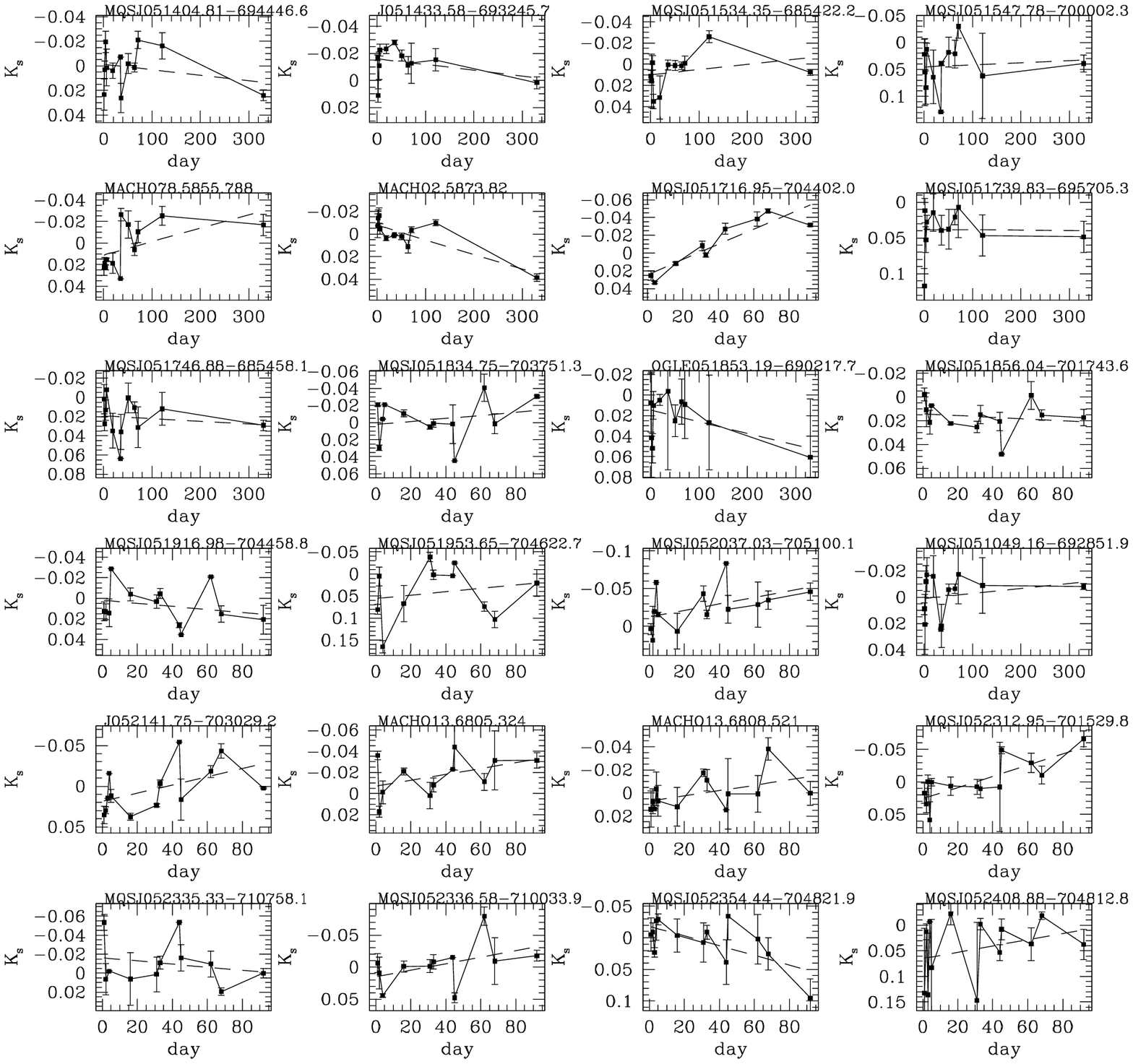}}
\caption{continued.}
\label{qsocurves4}
\end{figure*}

\setcounter{figure}{0}
\begin{figure*}
\resizebox{\hsize}{!}{\includegraphics{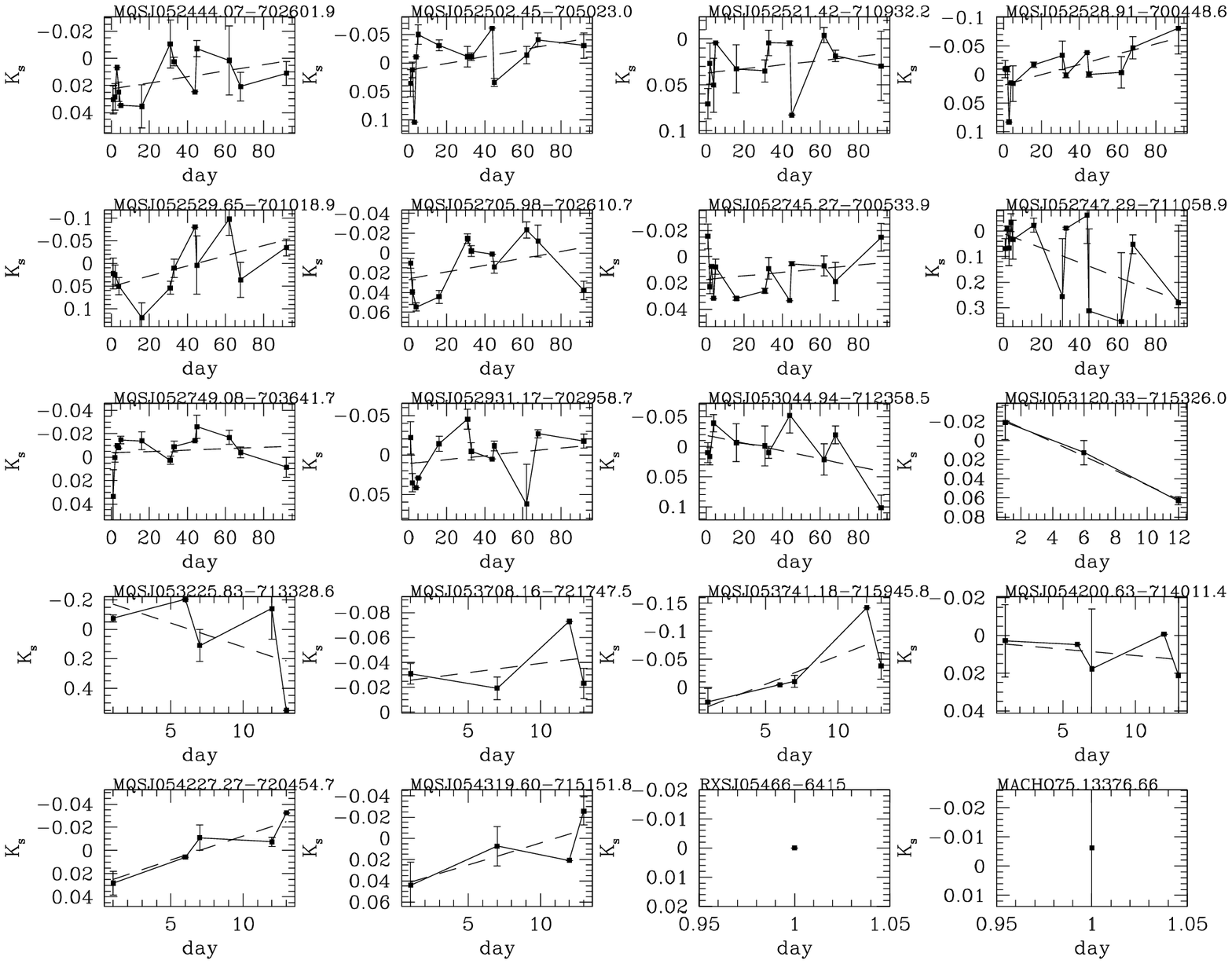}}
\caption{continued.}
\label{qsocurves5}
\end{figure*}

\end{appendix}

\end{document}